\documentclass[]{aa}

\newcommand{\apropto}{\;
  \raise0.3ex\hbox{$\propto$\kern-0.75em\raise-1.1ex\hbox{$\sim$
  }}\;\hskip-2pt }
\usepackage[dvips]{graphicx}

\newcommand{\lta}{\;
  \raise0.3ex\hbox{$<$\kern-0.75em\raise-1.1ex\hbox{$\sim$
  }}\;\hskip-2pt }
\newcommand{\gta}{\;
  \raise0.3ex\hbox{$>$\kern-0.75em\raise-1.1ex\hbox{$\sim$
  }}\;\hskip-2pt }

\begin{document}
\title{Polar branches of stellar activity waves: dynamo models and observations
}

   \author{D.~Moss\inst{1} \and D.~Sokoloff\inst{2} \and A.~F.~Lanza\inst{3}}

   \offprints{D.Moss}

   \institute{School of Mathematics, University of Manchester, Oxford Road,
Manchester, M13 9PL, UK \and
Department of Physics, Moscow University, 119992 Moscow, Russia
\and
 INAF-Osservatorio Astrofisico di Catania, Via S. Sofia, 78,
95123, Catania, Italy}

   \date{Received ..... ; accepted .....}

\abstract{Stellar activity data provide evidence of  activity wave
branches propagating polewards rather than equatorwards (the solar
case). This evidence is especially pronounced for the well-observed
subgiant HR~1099. Stellar dynamo theory allows polewards propagating
dynamo waves for certain governing parameters.} {We try to unite
observations and theory.} {Taking into account the preliminary
stage both of observations of polar activity branches and of the
determination of the governing parameters for stellar dynamos, we
restrict our investigation to the simplest mean-field dynamo models,
while recognizing more modern approaches to be an essential
development.}{We suggest a crude preliminary systematization of
the reported cases of polar activity branches. Then we present
results of dynamo model simulations which contain magnetic
structures with polar dynamo waves, and identify the models which
look most promising for explaining the latitudinal
distribution of spots in dwarf stars. 
Those models require  specific features of
stellar rotation laws, and so observations of polar activity branches
may constrain internal stellar rotation. Specifically, we find it unlikely that a pronounced poleward branch 
can be associated with a solar-like internal rotation profile, while it can be 
more readily reproduced in the case of a cylindrical rotation law 
appropriate for fast rotators.
We stress the case of the subgiant component of the  active close binary
 HR~1099 which, being best investigated, presents the most severe problems for a dynamo interpretation.  Our best
 model requires dynamo action in two layers separated in radius. This interpretation requires
some change in the paradigm of stellar magnetic studies, as it explains surface manifestations in a subgiant as a
joint effect of shallow and deep layers of the stellar convective
zone, rather than of a surface magnetic field only,
as appears to be the case in dwarf stars.
}{Observations of polar activity branches provide
valuable information for understanding stellar activity
mechanisms and internal rotation, and thus deserve intensive
observational and theoretical investigation. Current
stellar dynamo theory seems sufficiently robust to accommodate the phenomenology.}


\titlerunning{Polar branches of stellar activity waves}
\authorrunning{Moss et al.}

\maketitle

\section{Introduction}
\label{intro}

 It is widely accepted that the solar activity cycle is
more than just a quasiperiodic variation of sunspot number,
being rather an activity wave that propagates from mid solar latitudes
towards the solar equator. A solar dynamo based on the joint effects
of differential rotation and mirror-asymmetric convective motions in
the form of the so-called $\alpha$-effect (possibly together with meridional
circulation) is considered to be the underlying mechanism for the
 propagation of activity waves. Indeed, this mechanism gives an
equatorwards propagating wave of large-scale magnetic field for a
suitable choice of the parameters governing dynamo action. It is natural
to expect that such a phenomenon will appear in  a variety of stars with
convective envelopes, and we might thus be led to expect equatorwards waves of
stellar activity.

In fact cyclic activity is known now for many stars of various
spectral types, e.g.  \cite{Baliunasetal95}, \cite{Olahetal09}.
Clarification of the spatial configuration of the assumed activity
wave is a much more delicate undertaking. However contemporary
astronomy possesses a range of tools, such as the technique of
Doppler Imaging (hereafter DI), with which to investigate the problem. A
comprehensive investigation of the problem still remains a desirable
milestone for stellar astronomy; however some early results are
already available, e.g. \cite{BerdyuginaHenry07},
\cite{Katsovaetal10}. The point here is that at least some stars
exhibit an activity wave that propagates polewards.

For instance, the K-type subgiant component of the RS~CVn system
\object{HR~1099} has been extensively studied through DI by, e.g., \cite{Vogtetal99}, and shows  migration of spots from
mid-latitudes towards the rotational poles on a timescale of a few
years. Indirect evidence of the same phenomenon is found for several
late-type main-sequence  stars and young solar analogues, from
chromospheric line flux monitoring or photometric optical
monitoring, respectively (see Sect.~\ref{observations}).

In mean-field dynamo models, the direction of migration of the large-scale
magnetic field features depends in
principle on two key factors -- the sign of the $\alpha$ coefficient in the
relevant hemisphere and the
radial gradient of angular velocity.

The situation is however not so straightforward. The point is that
in addition to the equatorwards branch demonstrated by sunspots, the
solar activity displays a relatively weak polewards branch, seen in
some other tracers, e.g. polar faculae -- \cite{MS89}. It looks
implausible a priori that such weak additional branches could be
responsible for stellar polewards branches, however the possibility should
be recognized.

The aim of this paper is to investigate how the recent and current
observational data concerning polewards branches of stellar activity
might be connected with ideas from stellar dynamo theory. We
appreciate that the observational situation after these pioneering
results still remains quite uncertain, and so we study just the most
traditional forms of stellar dynamos, i.e. mean-field dynamos based
on differential rotation and $\alpha$-effect with simple algebraic
quenching.

More recent ideas in solar dynamo theory, such as flux transport
dynamos based on meridional circulation, { e.g. \cite{BD95}, \cite{CSD95},} \cite{DG06}, or
dynamical schemes of dynamo saturation, e.g. \cite{Ketal03},
\cite{SB04}, are { certainly} likely to be important. We believe
however that a simple initial approach is desirable and therefore we
consider a classic dynamo wave model with a simple nonlinearity as
the basic model in our research. In this model, the activity wave propagation  is primarily associated with the joint action of
differential rotation and $\alpha$-effect rather than any effects of
 meridional circulation. It appears however that some
observational features of polewards propagation are difficult to
reproduce with this simple model. Therefore, we also investigate briefly some effects of meridional circulation.

{ We also note the possible role of low order dynamo models in 
elucidating some aspects of stellar magnetic behaviour, e.g. \cite{WS06},
but here we concentrate on models that we feel are more directly interpretable physically.
}

Previous theoretical investigations of stellar dynamos have focussed on reproducing the dependence of activity cycle periods on stellar parameters, in particular  the rotation period (see, e.g., 
\cite{noyesetal84b}, \cite{ossendrijver97}, \cite{jouveetal10} and references therein) or on explaining   high-latitude or polar spots that are not observed in the Sun  
(e.g., \cite{granzeretal00}, \cite{Holzwarthetal06}, \cite{isiketal07})). 
We present here for the first time a tentative systematization of the stellar 
butterfly diagrams that are now
emerging from the  observations, 
and try to explain the different behaviours by means 
of a simple mean-field dynamo model. 

{ We stress that our primary aim is not to produce `definitive' mean-field
 models for any of the observed behaviours. Rather we attempt to illustrate 
the degree of uncertainty inherent in mean-field parametrizations, and
also to show that many behaviours are reproducible by such models. Whilst the 
fundamental shortcomings of mean-field theory have attracted much interest, 
rather less attention has been given to, for example, investigating differences in 
behaviour caused by modest changes to parametrizations. This is, 
of course, two-edged. It reduces any predictive power of mean-field
modelling, but also illustrates the possibility of explaining non-standard
behaviours by more unusual regimes.}

\section{Butterfly diagrams and polewards activity waves for solar-like stars}
\label{observations}


We  introduce some  tentative systematization
 of the information concerning the migration of activity patterns as derived
from available observations. Distinct cases  are summarized in
Table~\ref{observations_tab}, which lists also our dynamo models
with features resembling those observed (see
Sect.~\ref{comparison}), and are briefly described below.
They are listed in order of increasing Rossby number ($Ro$), that is the ratio of the rotation
period of the star to the convective turnover time at the base of
the convection zone; $Ro$ can  be roughly related to the dynamo
number in the bulk of the convection zone (cf. \cite{Noyesetal84a}).

\begin{description}

\item[Case I:] Among the RS~CVn systems, the K-type component of
HR~1099 has a record of DI maps extending over
about twenty years with simultaneous coverage in wide-band optical
photometry (cf., e.g., \cite{S09}). \cite{BerdyuginaHenry07}, extending previous work by
\cite{Lanzaetal06}, built maps of the distribution of  starspots  on the active
K-type subgiant.  Two main active regions were found, one migrating from
high latitudes ($\approx 70^{\circ}$) towards mid-latitudes ($\approx
40^{\circ}$), the other from mid-latitudes ($\approx 40^{\circ}$)
towards high latitudes ($\approx 70^{\circ}$), these occurring
more-or-less simultaneously.

Several other RS~CVn binaries  show a general behaviour similar to that of
HR~1099, although their DI and photometric time series are less
extended or have a more limited simultaneous coverage. A
characteristic of the active components of RS~CVn binary systems is
the presence of a polar spot that persists with little modification
over timescales of decades, and which may be due to the advection of
magnetic flux to the polar region of the star by diffusion or
meridional flows, e.g. \cite{SchrijverTitle01,Mackayetal04,Holzwarthetal06}.
In contrast, starspots at intermediate and low latitudes
seem to follow a cyclic migration that
might be associated with an oscillating
dynamo, cf. \cite{S09}.

\item[Case II:] In young, rapidly rotating stars, such as \object{AB~Dor} ($P_{\rm rot}=0.51$
days) and \object{LQ~Hya} ($P_{\rm rot}=1.66$ days), DI has revealed
the simultaneous presence of spots at high and low latitudes, as
well as a polar spot which has not been observed in all seasons,
thus indicating a less persistent feature than in the case of the
 RS~CVn systems (e.g., \cite{Kovarietal04}).
 In AB~Dor, the spots at low and intermediate latitudes do not appear
 to migrate significantly (\cite{Jarvinenetal05, Jeffersetal07}),
 in contrast to the case of HR~1099 (case I above).

\item[Case III:]
Solar-like stars with a rotation period of about $5-40$ days have been studied  through the long-term
monitoring of their chromospheric flux variations, mainly in the
framework of the classic Mt.~Wilson H\&K project and its recent
extensions, e.g. \cite{Baliunasetal95}, \cite{Baliunasetal98},
\cite{HallLockwood04}. \cite{DonahueBaliunas94} report that 36
stars out of about 100 have several  determinations of their
rotation period extending over several seasons.
Among them, 21 show patterns of rotation that vary with time or with
the phase of the activity cycle. Specifically, 12 stars display a pattern
that resembles what would be expected from the solar butterfly
diagram, although in six of them the rotation period increases as
the cycle progresses, in contrast to the solar case. One of the best examples is
\object{HD~114710}, see \cite{DonahueBaliunas92}.

The dwarf stars showing anti-solar behaviour   seem unlikely to
possess an anti-solar pattern of surface differential rotation, i.e.
with the poles rotating faster than the equator, because this has
never been  found from DI observations (\cite{Barnesetal05}), or from
theoretical models of stellar rotation (e.g. \cite{Rudigeretal98}).
Therefore, a plausible explanation is that their active regions
migrate polewards rather than equatorwards, which is what  we
define as case III.

\item[Case IV:] There is another possibility  to produce the phenomenology described
in Case~III. If stellar activity is not confined to
latitudes close to the equator, but is well extended towards the
poles, in addition to a polewards dynamo wave there can be another
wave propagating from intermediate latitudes towards the equator.
Which of the two waves dominates the modulation of the stellar flux
depends on the inclination of the rotation axis with respect to the
line of sight. If the star is viewed approximately pole-on, the
polewards branch will dominate and the observed behaviour is anti-solar,
while if the inclination is low, the star shows a solar-like
behaviour because the low-latitude branch dominates,  as  suggested
by, e.g., \cite{MessinaGuinan03}.

\item[Cases V and VI:] Among the stars considered by \cite{DonahueBaliunas94}, there are
four  stars that seem to reverse the trend of rotation period variation at
mid-cycle; six stars that show two narrow, but well separated bands of
rotation, suggesting two stationary active latitude belts -- we define this as case V;
and, finally, stars that have hybrid patterns with one band showing a variation of the
rotation period versus the cycle phase, while the other remains
fairly constant -- we take this behaviour to be representative of
case VI. Stars with one or
two fixed rotation periods as determined from Ca II H\&K monitoring
might be characterized by a standing dynamo
wave, e.g. \cite{Baliunasetal06}.
\end{description}

Cases III, IV, V, and VI could be different manifestations of the same kind of activity
behaviour, which appear distinct, either because of a different
inclination of the stellar rotation axis which emphasizes either the
polar or the equatorial region of a star, and/or a different
intensity of the activity, or a phase shift between the polewards and
the equatorwards branches of the butterfly diagram during the
activity cycle. The available observations are still too limited
to arrive at any sound conclusion on this point.
Therefore, we prefer to consider all the
suggested behaviours  separately because each case
makes a specific requirement for the theoretical butterfly diagram.

\begin{table*}
\caption{A systematization of the cases where polewards migration
or other features different from the solar case are  deduced to explain
observations. EW means equatorwards propagation, PW polewards propagation.
If our investigations result in a model that appears
suitable to explain a case, we identify it in the third column.}
\begin{center}
\begin{tabular}{|r|l|l|}
\hline
Case  & Observational features/example  & Dynamo interpretation \\
\hline
 I &  PW migration from mid-latitude and EW migration  & \\
   & from high latitude (HR~1099)  & See Sect.~\ref{HR1099} \\
\hline
 II & Spot pattern extended in latitude &   \\
 & with high and low latitude spots, &  \\
 & but no definite migration during the cycle (AB~Dor) &  Model  24\\
 \hline
 III &  PW migration, i.e. the spot rotation period increases & \\
& as the cycle progresses, contrary to the solar case & Model 13c \\
 \hline
 IV & EW at low latitudes, PW at high latitudes & \\
  & (the latter can dominate when the star is viewed pole-on) & Model 8\\
 \hline
V & Two separated narrow activity bands & Possible after a further\\
& &  specialization of stellar hydrodynamics\\
 \hline
 VI& Migration + a standing pattern & Model 19\\
 \hline
\end{tabular}
\end{center}
\label{observations_tab}
\end{table*}

\section{Mean-field dynamos}

Now we address the problem from the other aspect and discuss how the
butterfly diagrams appear in an assortment of stellar
and solar-like dynamo models.
We discuss here the simplest cases from the viewpoint
of dynamo theory, i.e. standard mean-field dynamos, e.g. \cite{RH},
with conventional boundary conditions and numerical
implementation, e.g. \cite{br}.

We investigate solutions of the standard mean field dynamo equation:
\begin{equation}
\frac{\partial{\vec B}}{\partial t} = \nabla\times({\vec u}\times
{\vec B}+\alpha{\vec B}-\eta\nabla\times{\vec B}), \label{dyneq}
\end{equation}
where $\eta$ is the turbulent diffusivity and $\alpha$ represents
the usual isotropic alpha-effect. The velocity field ${\vec u} =
\Omega \varpi \hat{\phi} + {\vec u}_{\rm m}$, where $\Omega$ is the angular
velocity of rotation, $\varpi$ the distance from the rotation axis, $\hat{\phi}$
the unit vector in the azimuthal direction, and ${\vec u}_{\rm m}$ the
meridional flow. We restrict our investigation to
axisymmetric solutions and solve the dynamo problem in a spherical shell, $r_0\le
r\le 1$, where $r$ is the fractional radius. We adopt $r_0=0.64$ for many of the
models, but we also investigate models with a deeper dynamo region with
$r_0=0.2$.

When modelling stellar magnetic fields, it is necessary to ensure
that the field in the interior joins smoothly on to a force-free
field in the external, very low density, region. Splitting the
magnetic field into poloidal and toroidal parts, ${\bf B}={\bf
B}_P+{\bf B}_T$, the  Lorentz force can be written as
\begin{eqnarray}
{\bf L}&=&(\nabla\times{\bf B_P})\times {\bf B}_P+(\nabla\times{\bf B_T})\times {\bf B}_T \nonumber \\
&+&(\nabla\times{\bf B_T})\times {\bf B}_P+(\nabla\times{\bf
B_P})\times {\bf B}_T. \label{lorentz}
\end{eqnarray}
For an axisymmetric field, the last term is identically zero, the
first two are poloidal vectors and the third is toroidal. In the present
case ${\bf B}_T=B_\phi$.
The condition ${\bf L}={\bf 0}$ can be satisfied by setting ${\bf
B}_\phi=0$ and $\nabla\times{\bf B}_P={\bf 0}$; this provides the
boundary condition on the interior field applied at $r=1$.
At the lower boundary we use 'overshoot' boundary conditions, simulating
the decay of the field to zero over a skin depth $\delta$, in the form
$\partial g/\partial r=g/\delta$, where $g$ represents the azimuthal component
of the vector potential for the poloidal field, or the toroidal field. These
boundary conditions have been used before, and have been shown to have no
significant effect on the results, except to reduce some field gradients
near the base of the convective region. 

\begin{figure}
\begin{tabular}{ll}
\includegraphics[width=0.3\hsize]{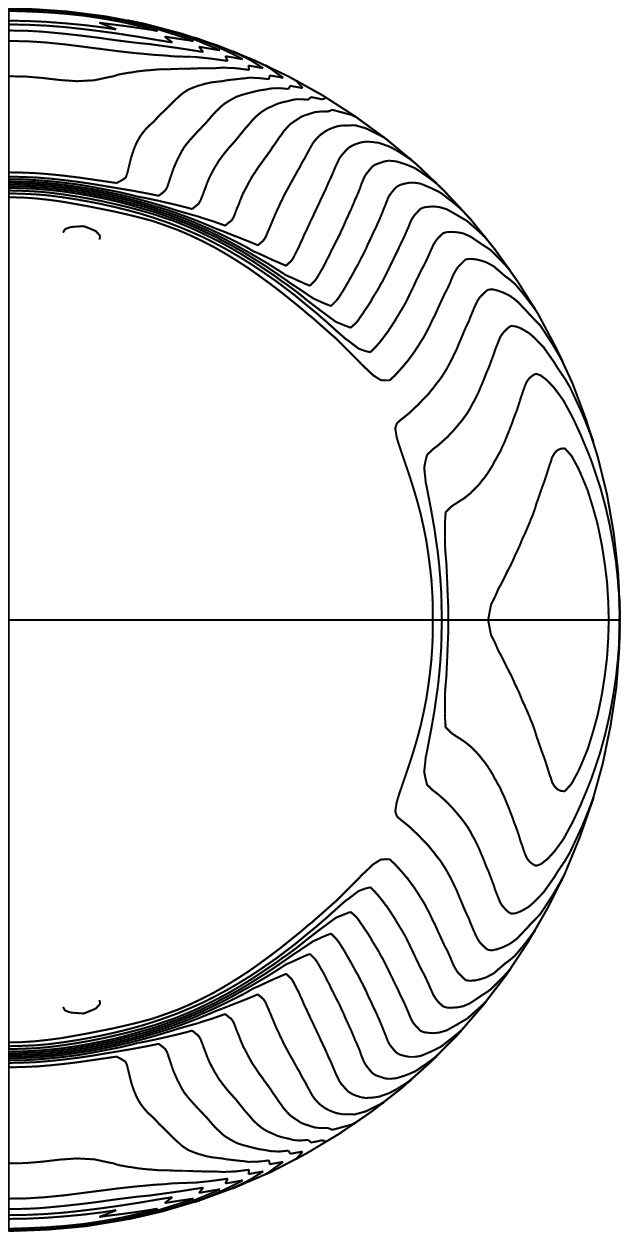} & \includegraphics[width=0.55\hsize]{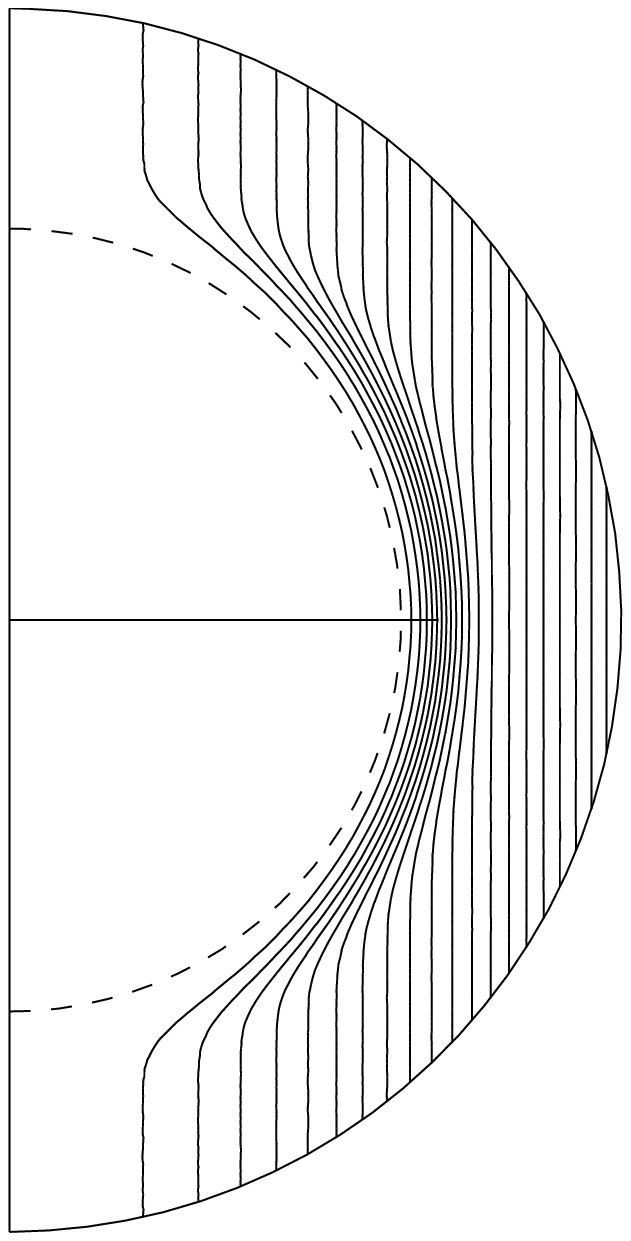} \\
\end{tabular}
 \caption{Equally spaced isorotation contours for the solar rotation law used
(left hand panel) and contours for our quasi-cylindrical law
for rapid rotators (right hand panel).}
 \label{rotation}
\end{figure}

Models with a  solar-like
rotation law, based on that derived from helioseismology,
and also models with a quasi-cylindrical rotation law appropriate to rapidly rotating lower mass dwarfs,
are studied  (Fig.~\ref{rotation}), with
a variety of choices for $\alpha(r,\theta)=C_\alpha
f_1(r)f_2(\theta)/[1+({B}/B_{0})^2]$, i.e. a
naive $\alpha$-quenching nonlinearity is used, { where $\theta$ is the colatitude measured from the North pole and $B_{0}$ a reference magnetic field (see below).}

Different physical mechanisms can co-operate to produce the
$\alpha$-effect and theoretical or observational (e.g. \cite{Zhang})
knowledge of its radial and colatitudinal
distributions, here expressed through the functions $f_{1}(r)$ and $f_{2}(\theta)$, is quite preliminary. Therefore, we adopt only simple
parametrizations and explore a number of
options to investigate the sensitivity of butterfly diagrams to the
underlying assumptions.
{The turbulent diffusivity $\eta$ is uniform in the outer part of the convection zone ($r\geq 0.8$), but decreases linearly to one half that value in the domain $r \leq 0.7$. 
{ Below the CZ proper we expect an overshoot region,
where the turbulent intensity, corresponding to the turbulent
resistivity, is further reduced. In \cite{mossbrooke00} (which uses
the Malkus-Proctor feedback onto the differential rotation as the sole
nonlinearity, as opposed to the algebraic alpha-quenching here),
only token recognition of this effect was made for computational reasons.
We have followed the same procedure here, recognizing that the gradient in
turbulent diffusivity should be larger.  
Another way of regarding this is to consider the model as having a 
rather deeper CZ, extending to radius $r_0$. From this viewpoint, 
the lower boundary condition, that allows a penetration of the field with a skin depth of the order of $\delta$, corresponds to the effect of a strongly reduced diffusivity immediately below the boundary. }

We make the dynamo equation non-dimensional in terms of the stellar radius $R$, the diffusion time $R^{2}/\eta_{0}$, where $\eta_{0}$ is the maximum turbulent diffusivity, and the magnetic field  $B_{0}$ defined as in Sect.~3 of \cite{mossbrooke00}. Thus 
we introduce the standard dynamo numbers, $C_\alpha=\alpha_0R/\eta_0, C_\omega=\Omega_0R^2/\eta_0$,
where $\alpha_0$ is a typical value of $\alpha$ and $\Omega_0$ is
the maximum value of the angular velocity.} In the
$\alpha\omega$ approximation, we can define the combined dynamo number
$D=C_\alpha C_\omega$,  and this remains a useful quantity even in $\alpha^2\omega$ models.

To integrate the dynamo equation, we use the code   described in \cite{mossbrooke00},
which uses a Runge-Kutta integrator over a standard mesh with 61 points
over $r_0\le r\le 1$, and 101 points over $0\le \theta \le \pi$,
equally spaced. The results are described in Sect.~\ref{2d_dynamos}.

A useful simplification of the general axisymmetric mean-field dynamo equation
(\ref{dyneq}), known as the Parker migratory dynamo, is considered in Sect.~\ref{1D} and is
written in a standard non-dimensional form as:
\begin{equation}
\label{eq1} {{\partial B} \over {\partial t}} = Df \sin \theta
{{\partial A} \over {\partial \theta}} + {{\partial^2 B} \over
{\partial \theta^2}} - \mu^2 B, \label{parkerB} \label{1D-eq1}
\end{equation}
\begin{equation}
\label{eq2} {{\partial A} \over {\partial t}} = \alpha B +
{{\partial^2 A} \over {\partial \theta^2}} - \mu^2 A, 
\label{parkerA}
\end{equation}
where { $f$ is a factor that allows for a latitudinal variation of the radially averaged angular velocity (see Sect.~\ref{1D})}, $\alpha$ is a nondimensional measure of the $\alpha$ effect,
$D=C_\alpha C_\omega$  is the dynamo number, $B$ denotes the toroidal
magnetic field and $A$ the toroidal component of the vector
potential for the poloidal field. Both latter quantities are
averaged in radial direction over the convective shell, { see
\cite{Baliunasetal06}.} In other words, here the explicit radial
dependence has been removed, and the terms involving $\mu^2$
represent radial diffusion in a spherical shell of thickness
approximately $\mu^{-1}$ of the outer radius of the shell -- e.g.
$\mu\approx 3$ is appropriate for the solar convection zone. 
{ We solve Eqs.~(\ref{eq1}) and (\ref{eq2}) in the domain $0 \leq \theta \leq \pi$ with 
$A=0$, $B=0$ at the boundaries. }

\subsection{Polar branches in the 1D Parker dynamo}
\label{1D}

We begin with a simple cartoon which explains the idea of stellar
dynamos, i.e. with the 1D Parker (1955) dynamo.

\begin{figure}
\includegraphics[width=0.92\hsize]{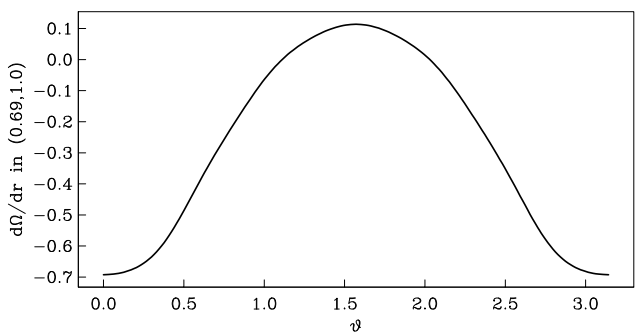}
\caption{Function $f (=<\frac{d\Omega}{dr}>)$ used in the Parker model
\protect(Sect.~\ref{1D}).}
 \label{rr}
\end{figure}

First of all,  we estimate the latitudinal variation of $\Omega$,
averaged over fractional radii (0.69,1) from a realistic solar rotation law
(Fig.~1, left hand panel). Normalized to the polar value of the
gradient, the modulation compared to that with no latitudinal
variation is modelled  by a function $f(\theta)$ (Fig.~\ref{rr}).
Then we ran the Parker dynamo with the usual dynamo number, $D=D_0$
say, replaced by $D=D_0 f(\theta)$, so $f(\theta)=1$ gives the
standard case. 

We take $\alpha=\cos\theta\sin^m\theta$, with
$m=0,2,4$.  The larger the value of $m$, 
the more concentrated around the equator the $\alpha$ effect. 
This simple parametrization has been used by other authors, 
e.g. \cite{rudigeretal03}, \cite{charbonneau10}, and we also adopt it here.
{ We  stress the inherent uncertainty in the spatial dependence of $\alpha$ and refer to, e.g.,
\cite{rudigerbrandenburg95} for some justification in the framework of mean field theory with a specific turbulence model. 
Specifically, they explore  an $\alpha$ effect dependence with $m=2$ that is suggested 
by the extension of their turbulence theory to third order terms 
in ${\vec \Omega} \cdot {\vec U} \propto \cos \theta$, 
where $\vec \Omega$ is the angular velocity vector of the star and 
$\vec U$ the vector of the gradient of the turbulent diffusivity that points 
in the radial direction. In principle a further extension of the theory to 
the fifth degree in ${\vec \Omega} \cdot {\vec U}$ would introduce terms 
proportional to $\sin^{4} \theta $ as we assume in our simple parametrization 
with $m=4$ (of course  we should then also consider expressions
for alpha containing a combination of these dependences, but we decided that
this was beyond the scope of this paper). Note also that the $\alpha$ effect may have a component arising 
from  magnetostrophic waves excited in the layers where the toroidal field 
is amplified and stored. This would lead to terms with $m>0$ in the 
parametrization of the latitudinal dependence of the $\alpha$ effect 
(see, e.g., \cite{Schmitt87}).}

{ We emphasize that calculations of $\alpha$ from turbulence models
are necessarily severely truncated, and stellar dynamos operate
in regimes remote from those in which such calculations are valid.
Also, if we consider the potentially more useful, but more problematical, 
method of obtaining  parametrizations of $\alpha$ from analysis of direct numerical simulations, it would be surprising if $m=0$ or $m=2$ or a combination was adequate. However such 
determinations are currently contentious, in spite of substantial progress,
from the early attempts (e.g. Brandenburg \& Sokoloff 2002) to the most
recent (e.g. Courvoisier \& Kim 2009; Brandenburg et al. 2010; Tobias, Dagon \&
Marston 2011). Realistically, a reliable determination for specific types
of stars remains a remote possibility. 

With these assumptions on the  rotation profile
and the alpha effect, we solve the 1D dynamo equations (\ref{1D-eq1}) and (\ref{parkerA}). }
With $D_0<0$ then, as expected, the activity wave propagates
equatorwards. The main effect is with the variation of $m$, larger
values of $m$ move the migration nearer to the equator.

 With $D_0>0$, $m=0$ then, again as expected, there is polewards
migration, centred on mid-latitudes, both with $f(\theta)=1$ and
with $f(\theta)$ nonuniform. When $f(\theta)$ is nonuniform, the
butterfly diagram is concentrated at high latitudes for modestly
supercritical $D_0$,  while  solutions become steady for more
supercritical values of $D_0$.

With $m=2$ and $f(\theta)=1$, again behaviour is much as expected,
but quite unexpectedly  with $f(\theta)$ as in Fig.~\ref{rr} and
$m=2,4,$ the solutions are steady for only marginally supercritical
values of $D_0$. These statements are all for solutions with dipole
parity, but steady solutions are also found  when quadrupole parity
is enforced.


Except when explicitly stated, all the above experiments were made
with a slightly supercritical $|D_0|$ and with a decay term which
mimics radial diffusion for the value of the ratio $\mu^{-1}=h/R=1/3$,
where $R$ is the stellar radius and $h$  the thickness of the convective zone, e.g.  \cite{Baliunasetal06}.
We deduce that, even with this simple dynamo model,
behaviour can be remarkably rich and
broad conclusions cannot be drawn without a quite careful exploration of
the parameter space and the form of $f(\theta)$.

\begin{figure*}
\begin{center}
\begin{tabular}{ll}
(a) \includegraphics[width=5cm]{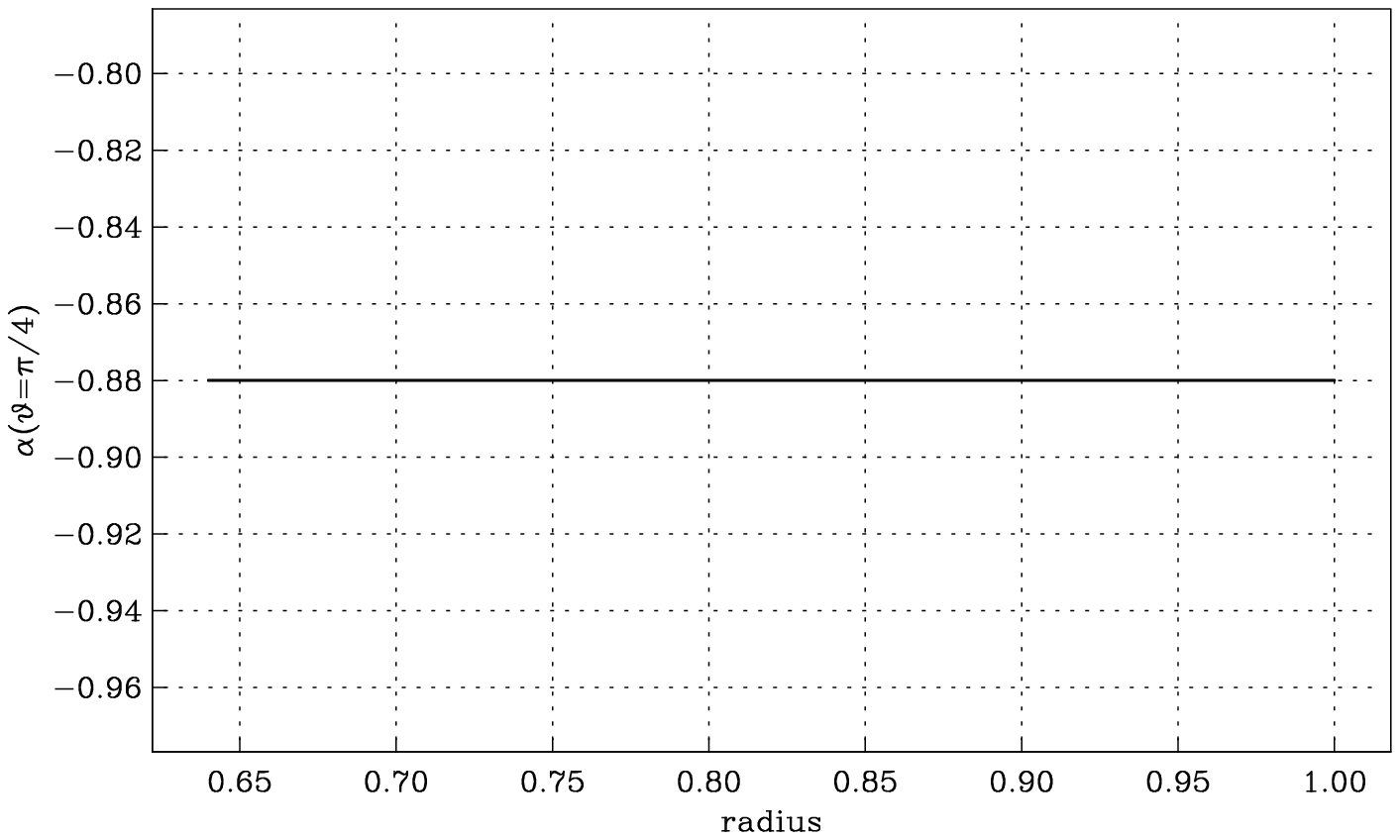} &
\includegraphics[width=5cm]{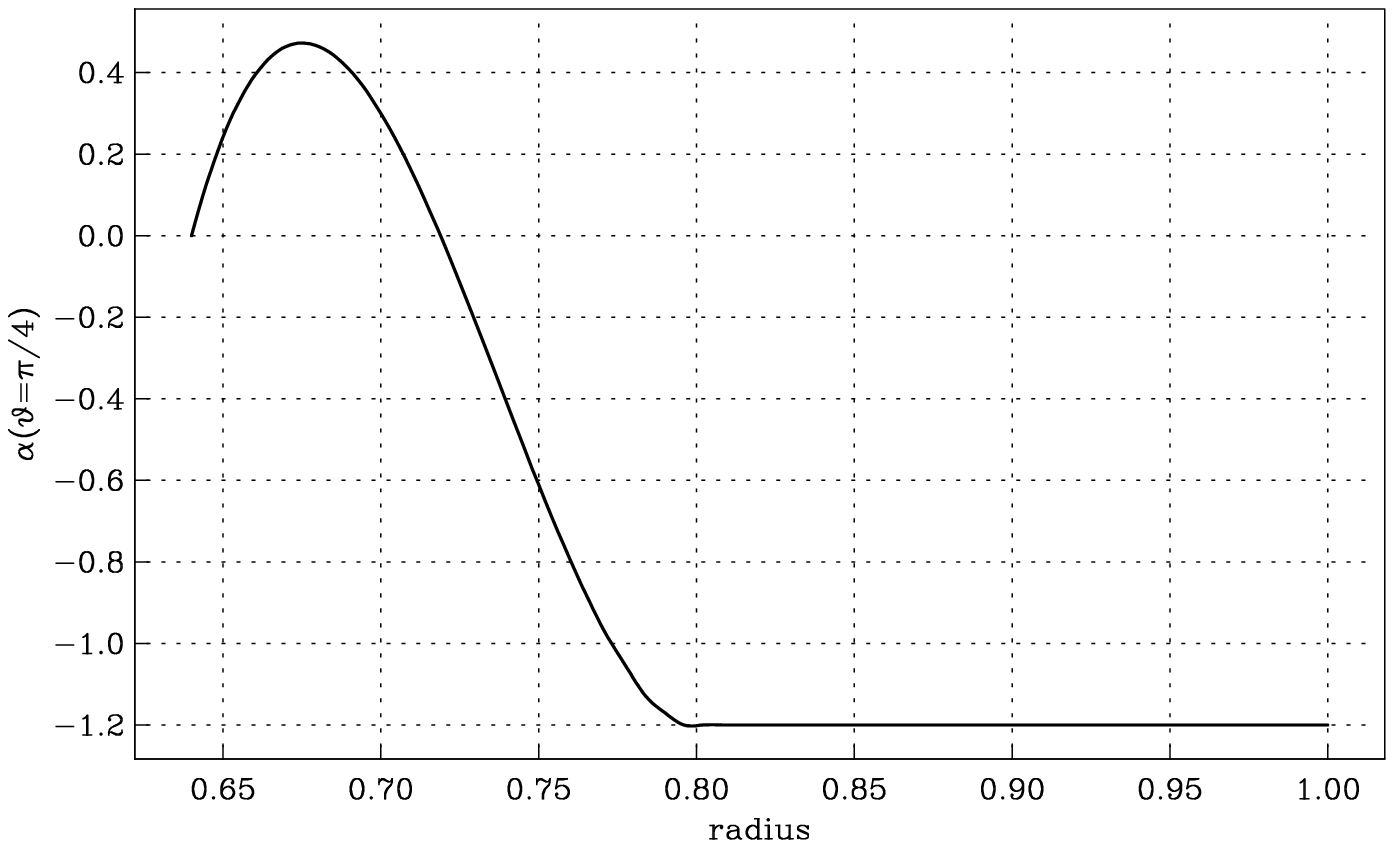}
\\
(b) \includegraphics[width=5cm]{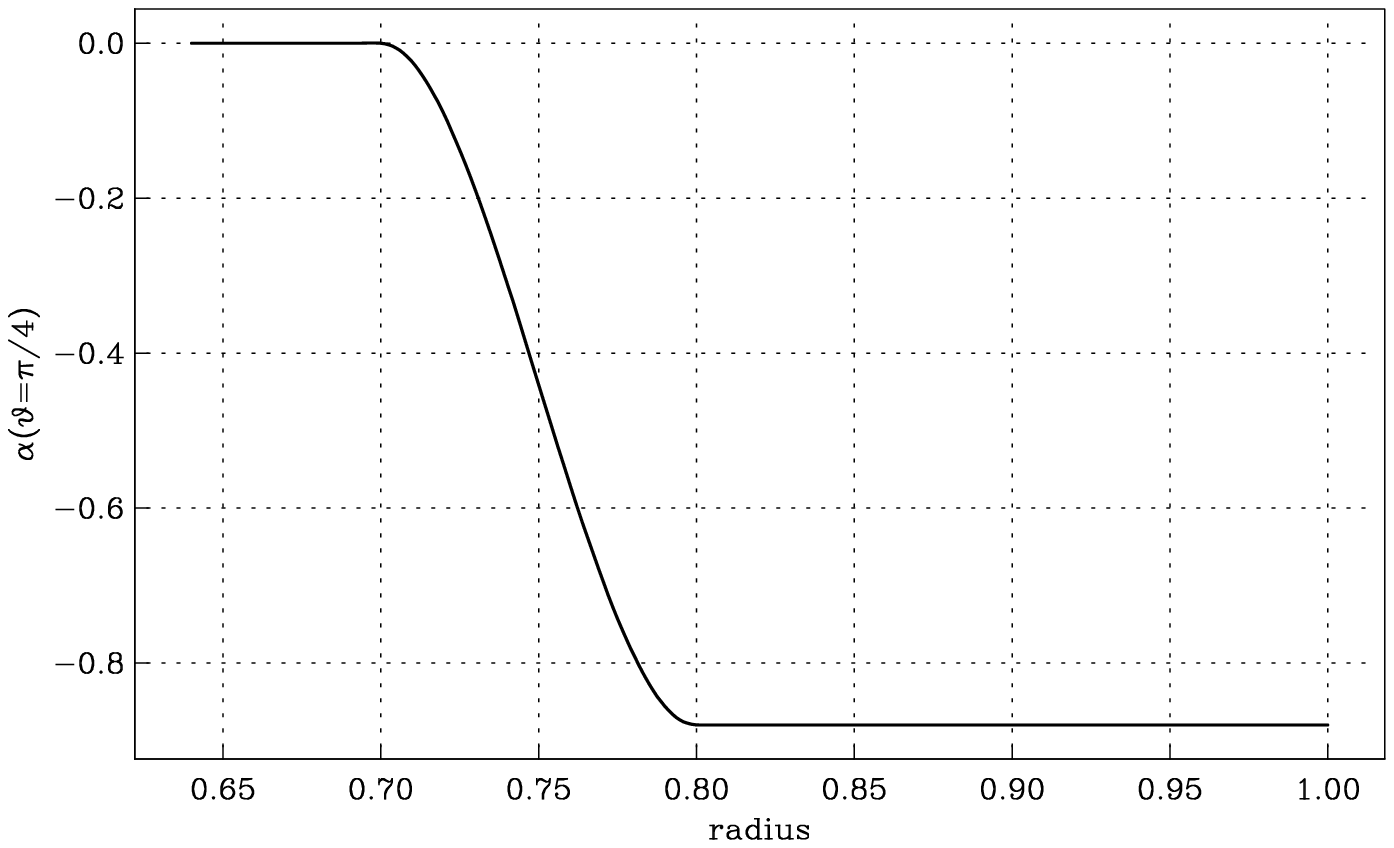} &
\includegraphics[width=5cm]{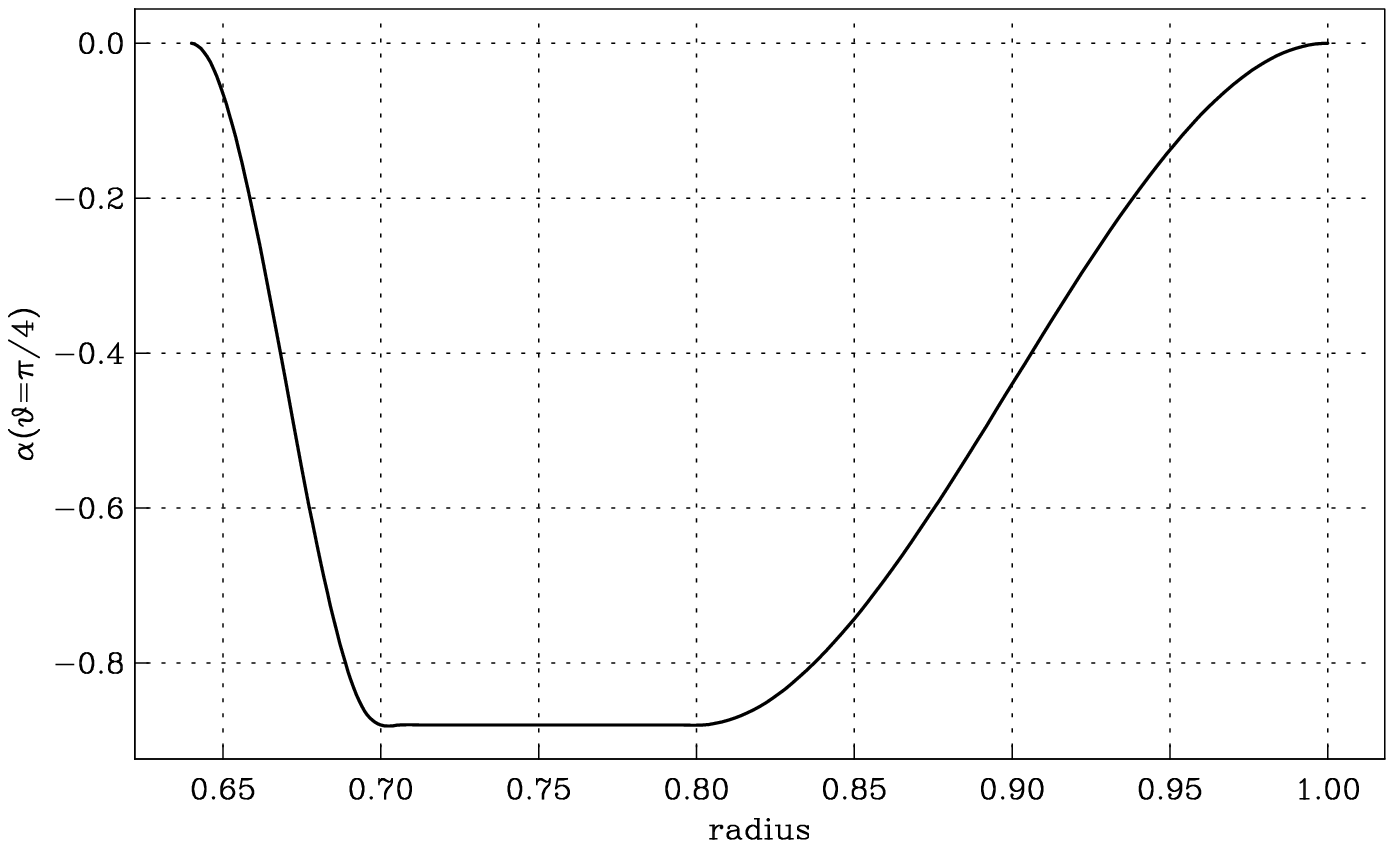}
\\
(c) \includegraphics[width=5cm]{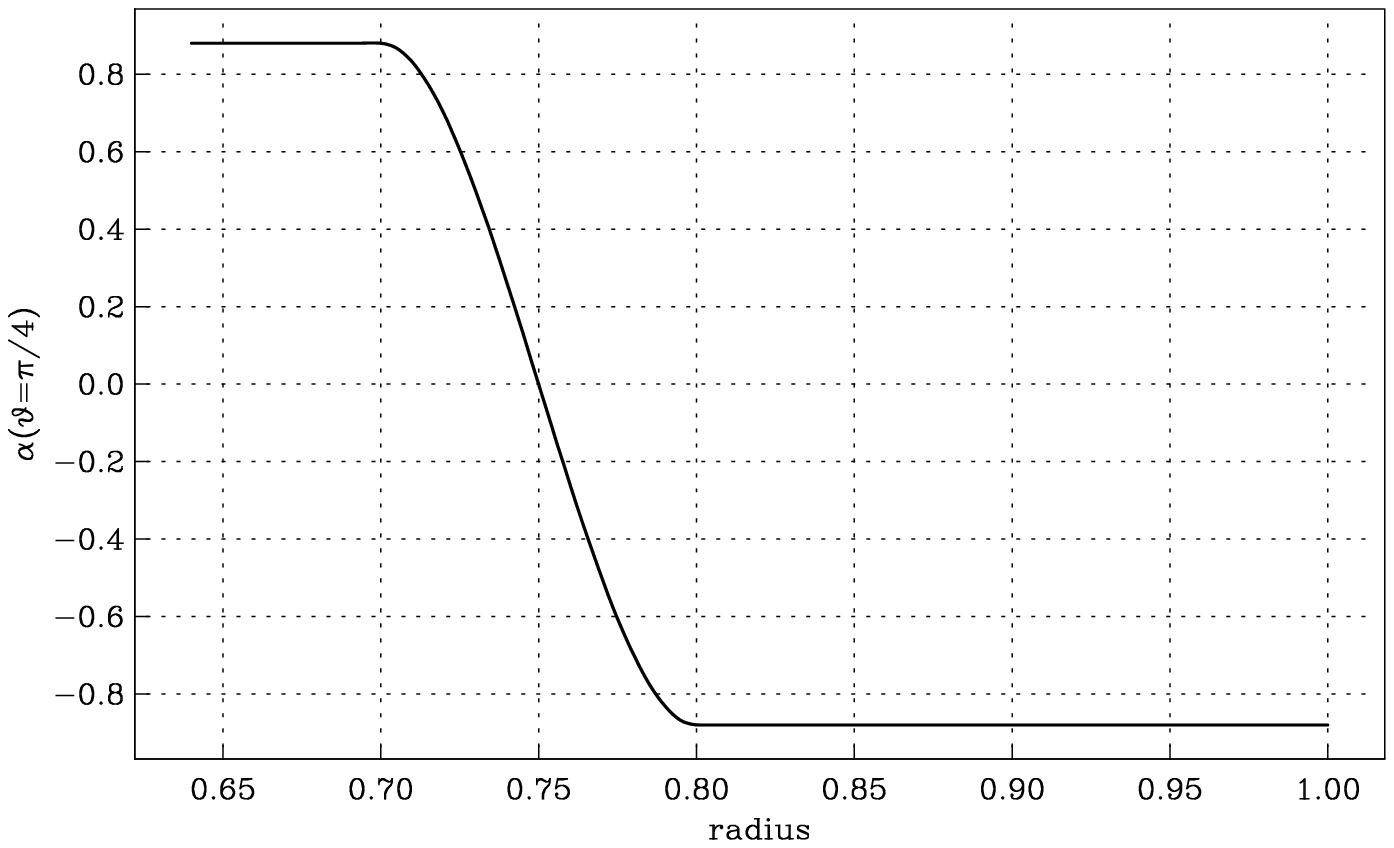} &
\includegraphics[width=5cm]{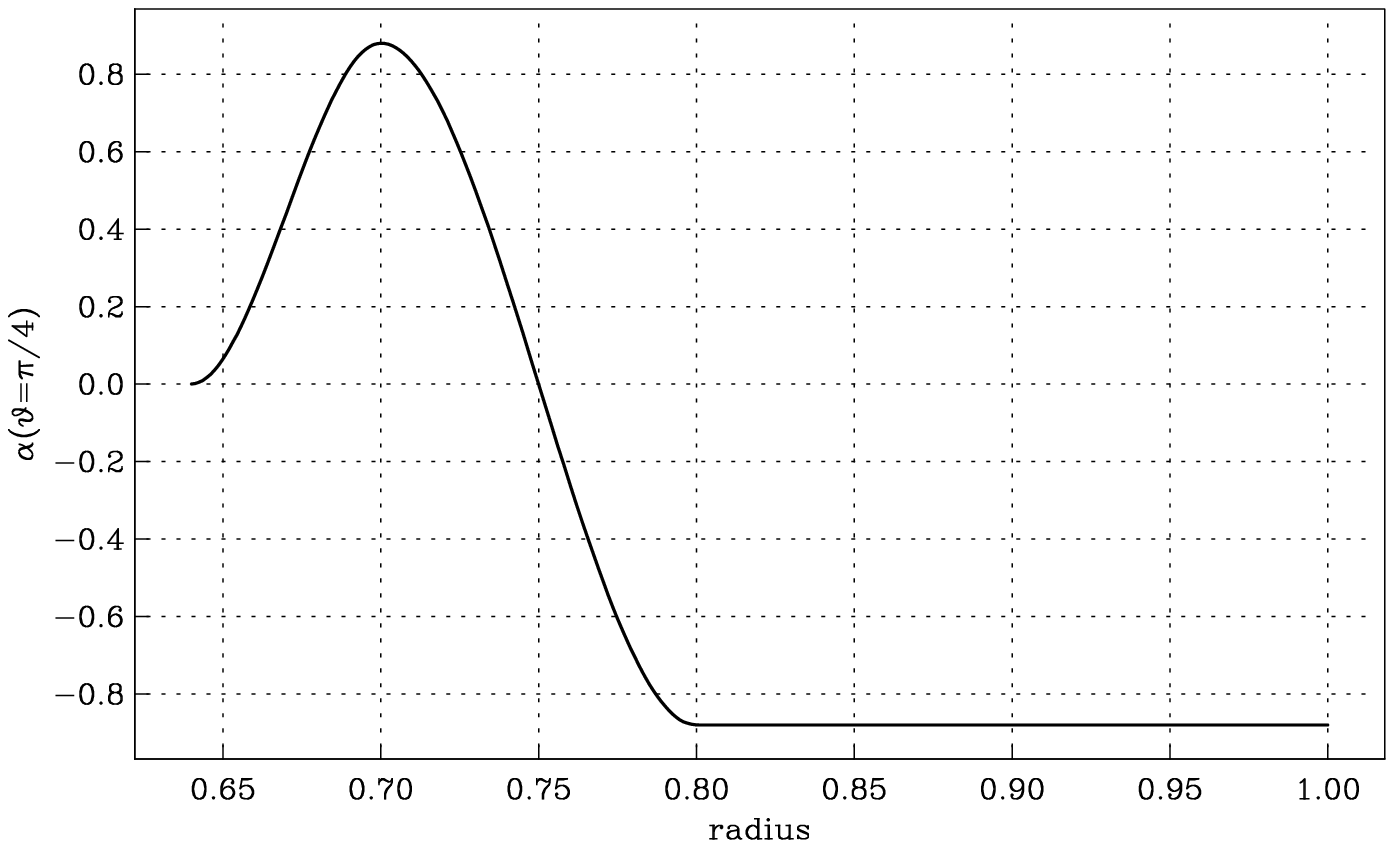}
\end{tabular}
\end{center}
\caption[]{\label{alp_prof} The function $f_1(r)$ (which determines
the radial dependence of the unquenched $\alpha$-term). Row (a),
$i_\alpha=0, 1$; row (b), $i_\alpha=2,3$; row (c), $i_\alpha=6, 7$.}
\end{figure*}

\subsection{2D dynamo models}
\label{2d_dynamos}

\subsubsection{Solar rotation law}

Now we turn our attention to 2D models and start by using a
solar-like rotation law  within fractional radii
$0.64\le r \le 1.0$, as shown in the left hand panel of
Fig.~\ref{rotation}. In $r\ge 0.7$ this is an interpolation on MDI
data, and is made to  match uniform rotation at the lower boundary
$r=r_0=0.64$.

A dynamo shell with a lower boundary at $r_0=0.64$ of the stellar radius
is a good assumption for G and K type main-sequence stars, but it is
not appropriate for a subgiant such as the K1IV active component of HR~1099.
Given the current uncertainty on the evolutionary stage of the star,
$r_{0}=0.2$ is a reasonable assumption (cf., e.g., \cite{lanza05}).
Because extensive numerical simulations to cover
realistic values for this (and many other) dynamo governing
parameters are obviously beyond the scope of the paper, we restricted
our investigation here to extrapolations that seem reasonable on
the basis of available knowledge.

Several forms of the functions $f_1(r), f_2(\theta)$ are  used:
$f_2(\theta)= \cos\theta \sin^m\theta$, with $m=0,2,4$, and $f_1(r)$
takes the forms shown in Fig.~\ref{alp_prof}, referred to below as
$i_\alpha=0,1,2,3,6,7$, as denoted in the caption. There are thus 18
possible forms of $\alpha(r,\theta)$.

A standard value $C_\omega=\Omega_{\rm eq}R^2/\eta_0 = 6\times 10^4$
was taken. The other dynamo number, $C_\alpha $, was given a
slightly supercritical value. 
 The results are summarized in Table~2.

\begin{figure*}
\begin{center}
\begin{tabular}{ll}
\includegraphics[width=7cm]{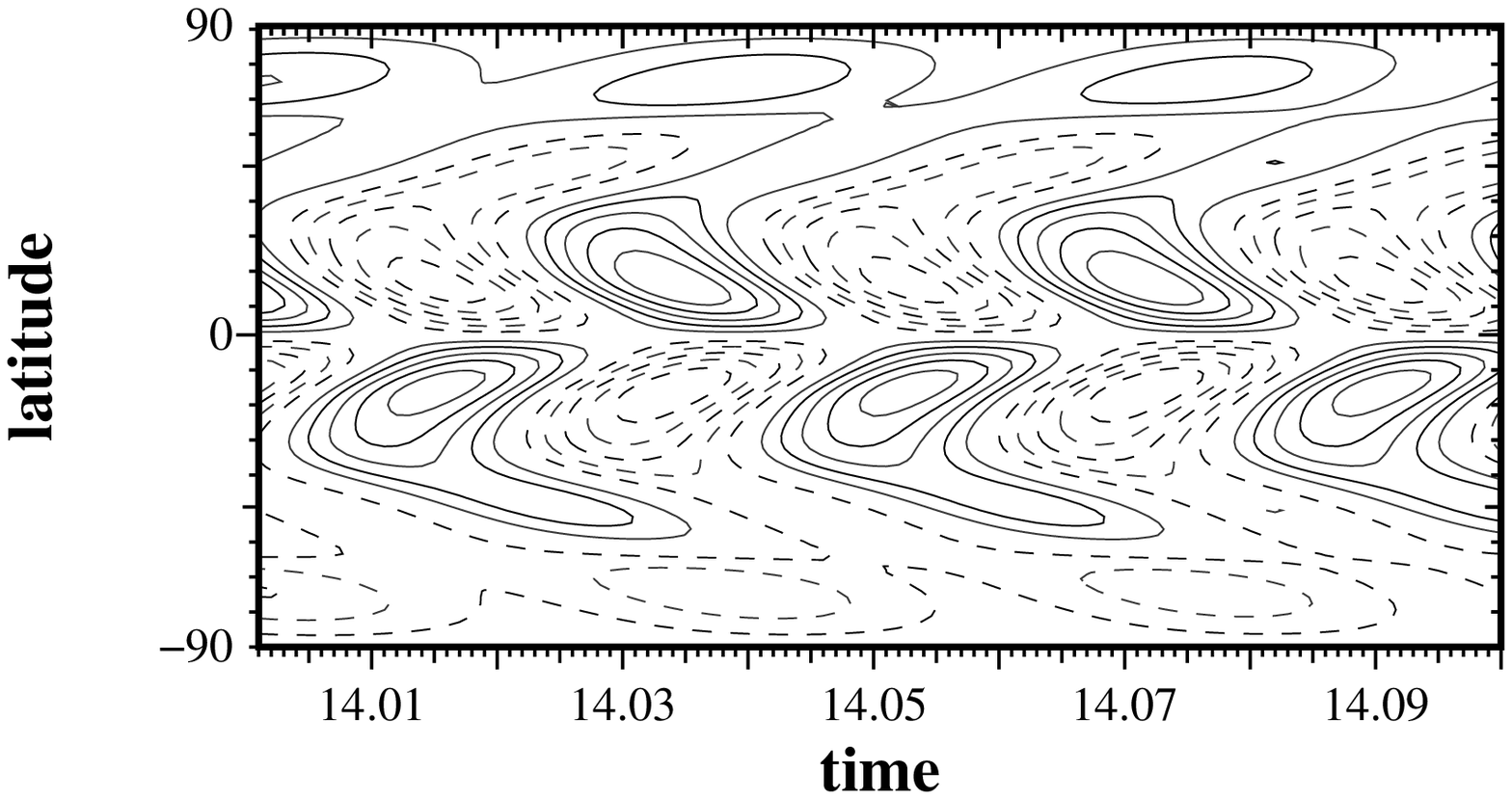} &
\includegraphics[width=7cm]{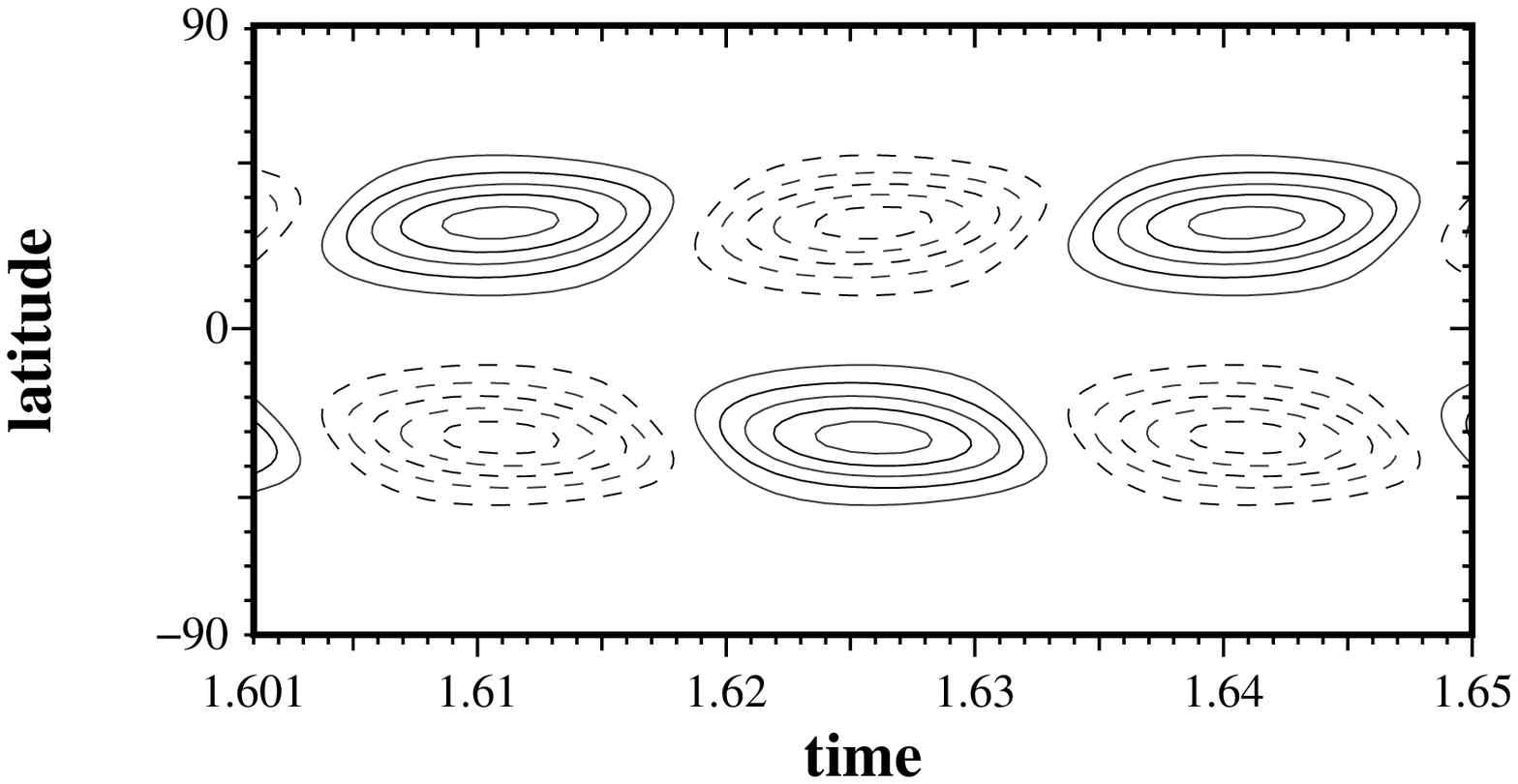} \cr
\includegraphics[width=7cm]{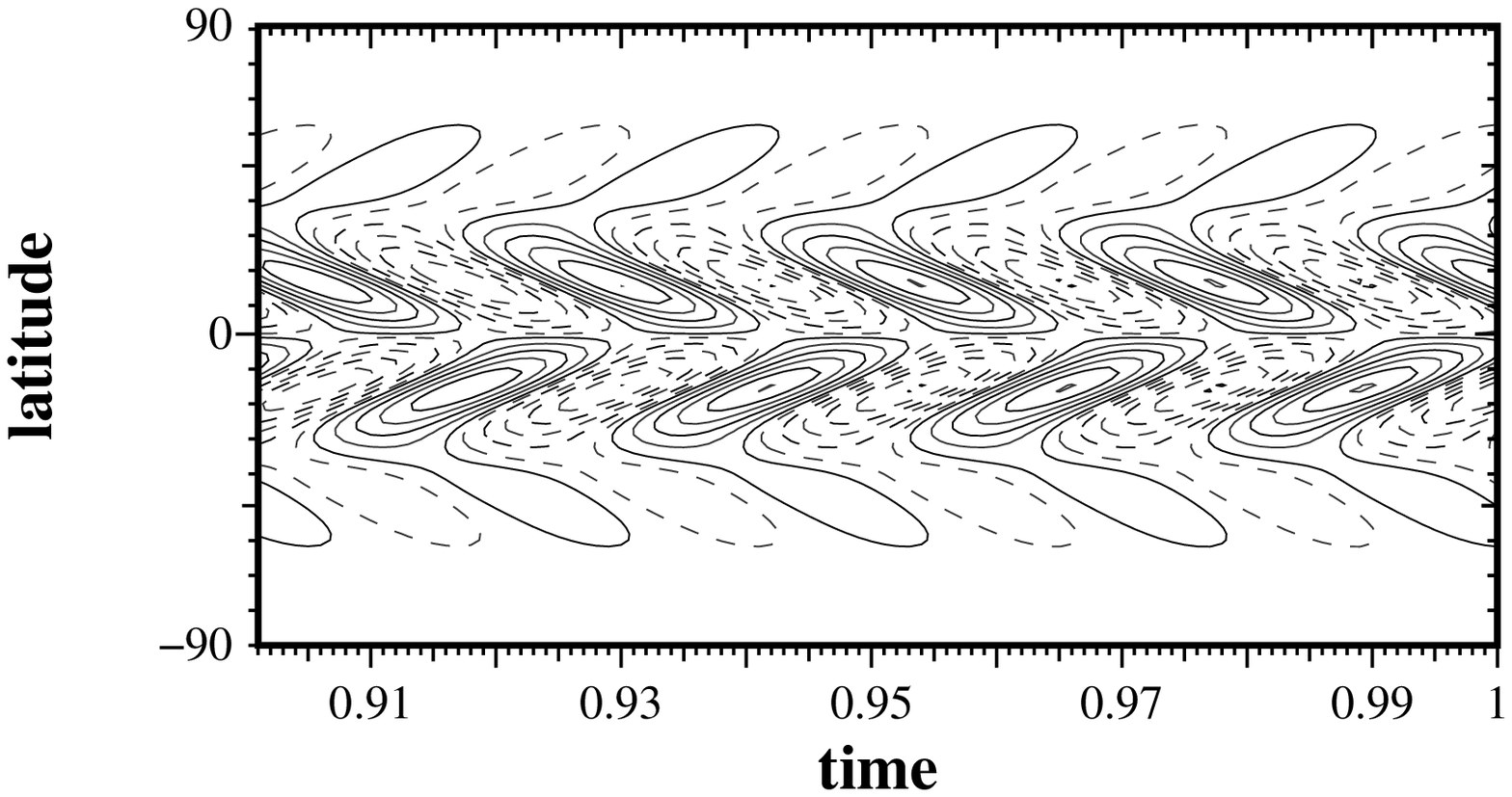} &
\includegraphics[width=7cm]{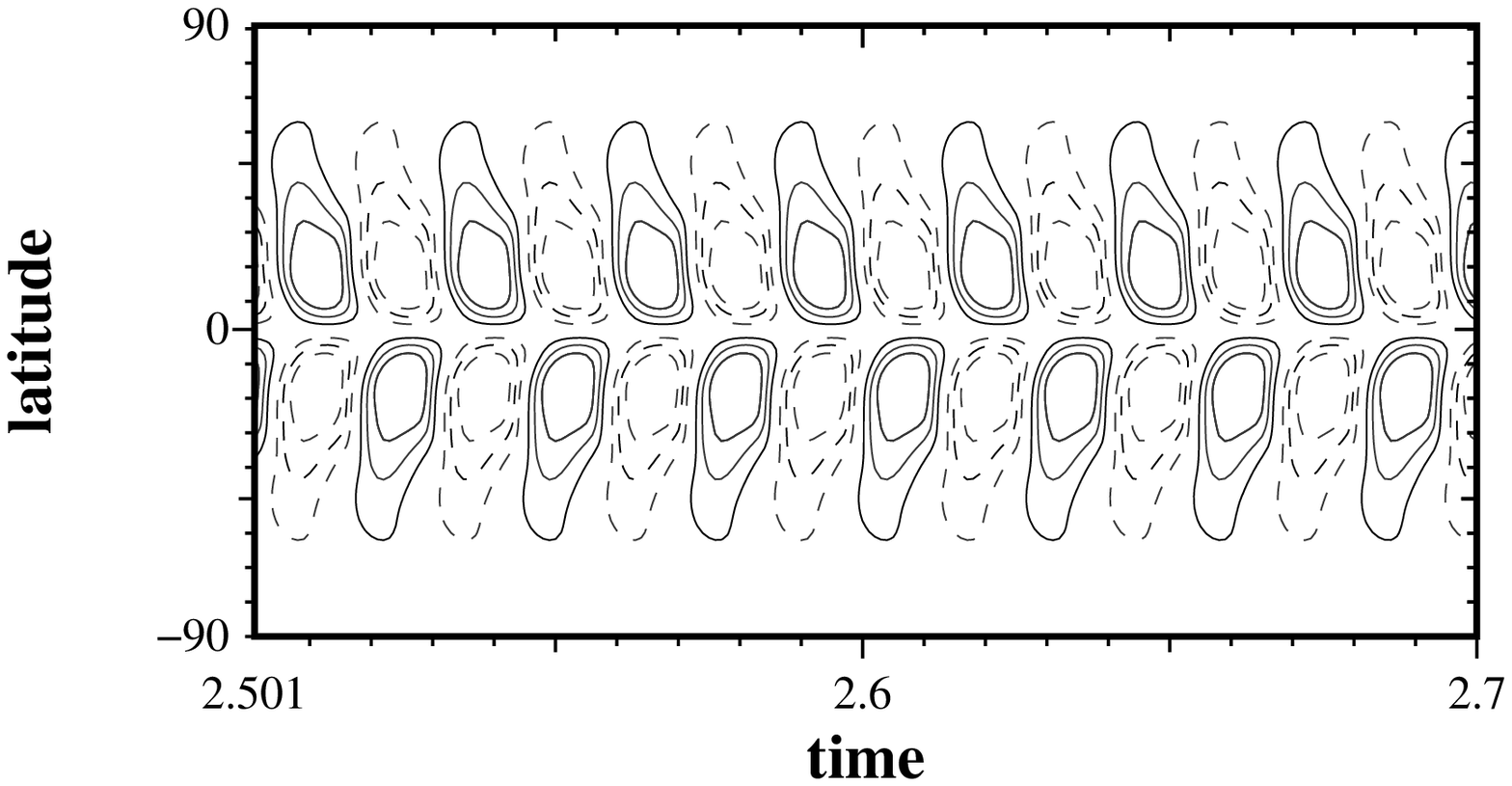}\cr
\includegraphics[width=7cm]{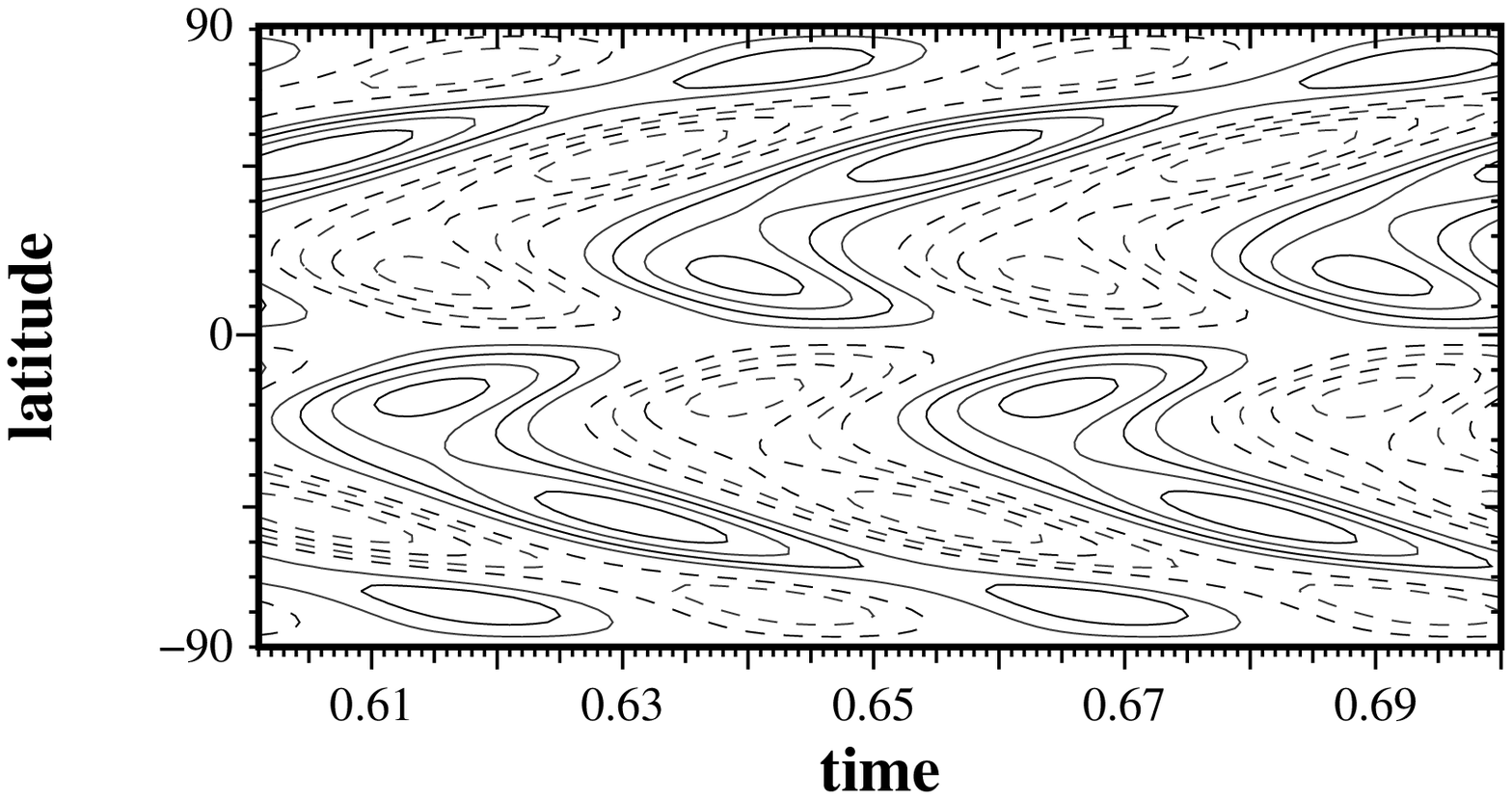} &
\includegraphics[width=7cm]{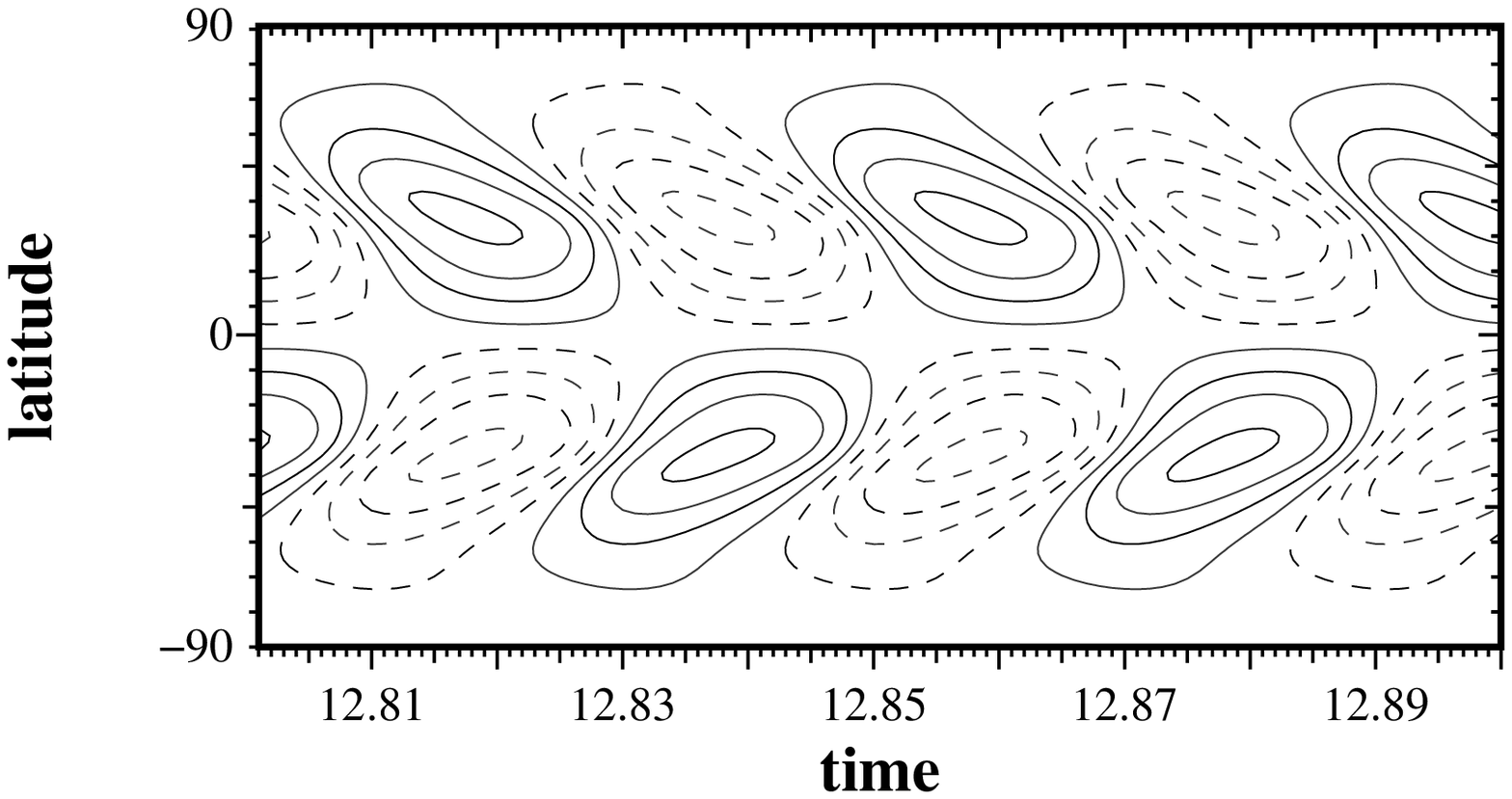}\cr
\includegraphics[width=7cm]{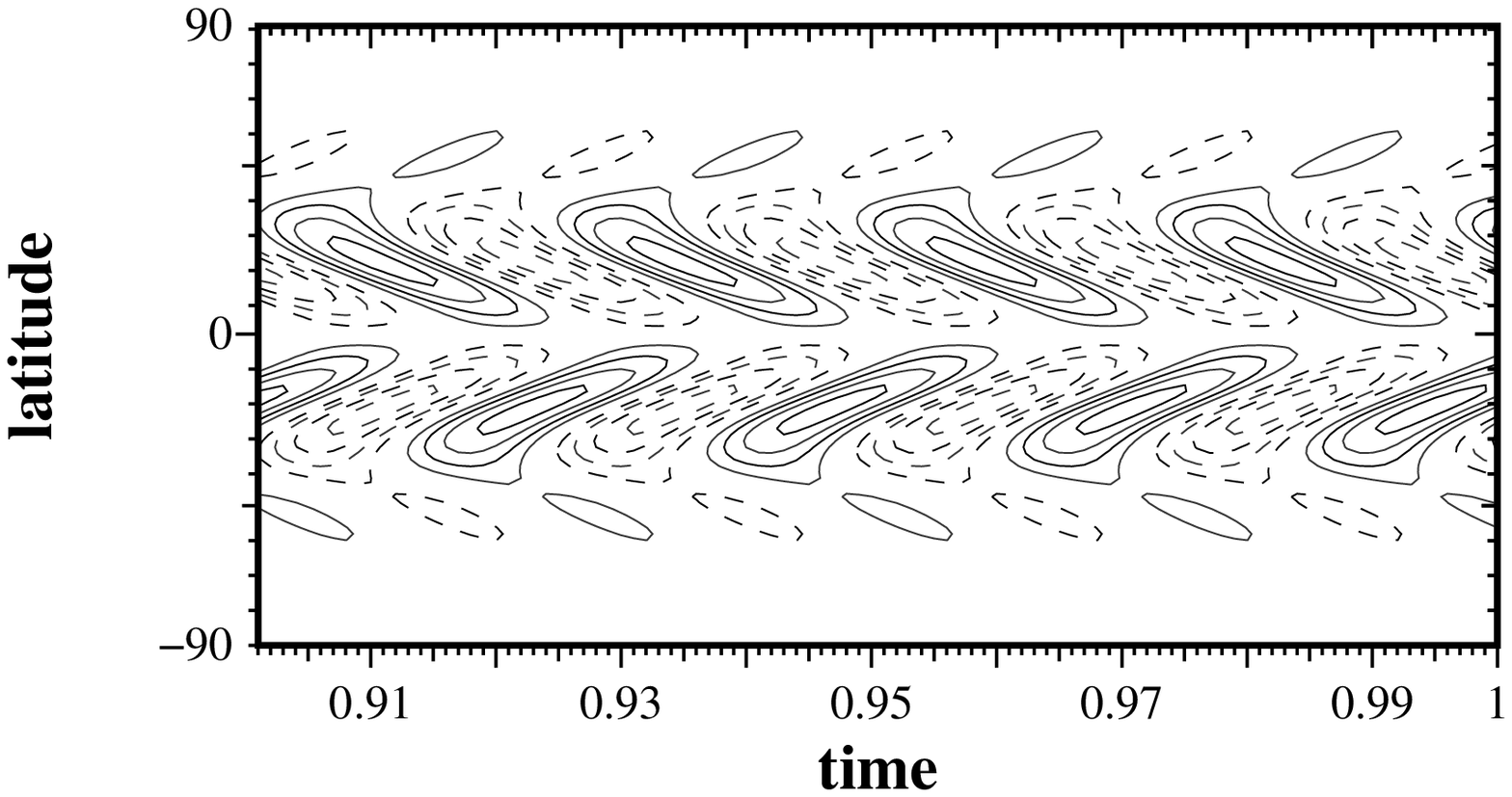} &
\includegraphics[width=7cm]{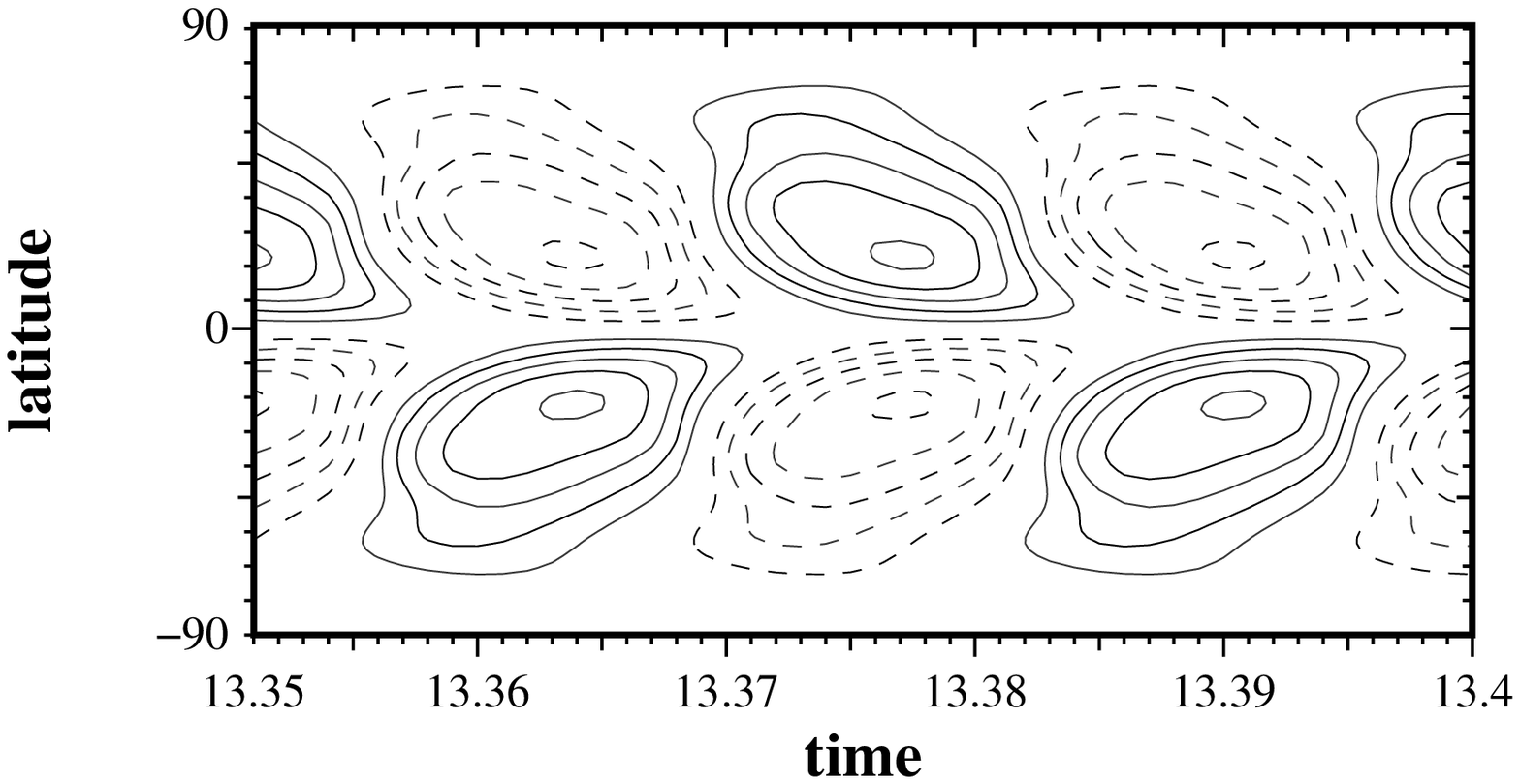}\cr
\includegraphics[width=7cm]{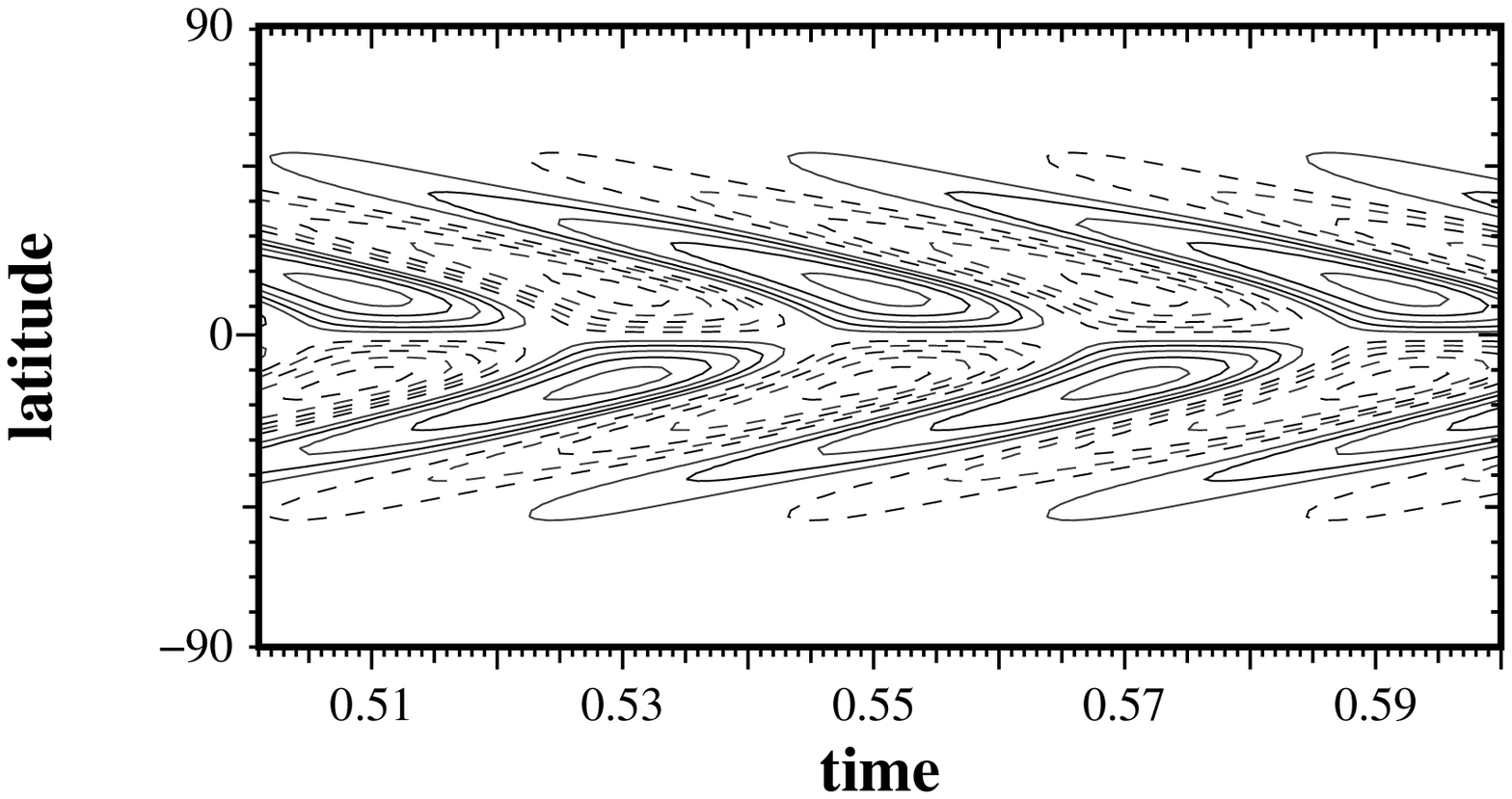} &
\includegraphics[width=7cm]{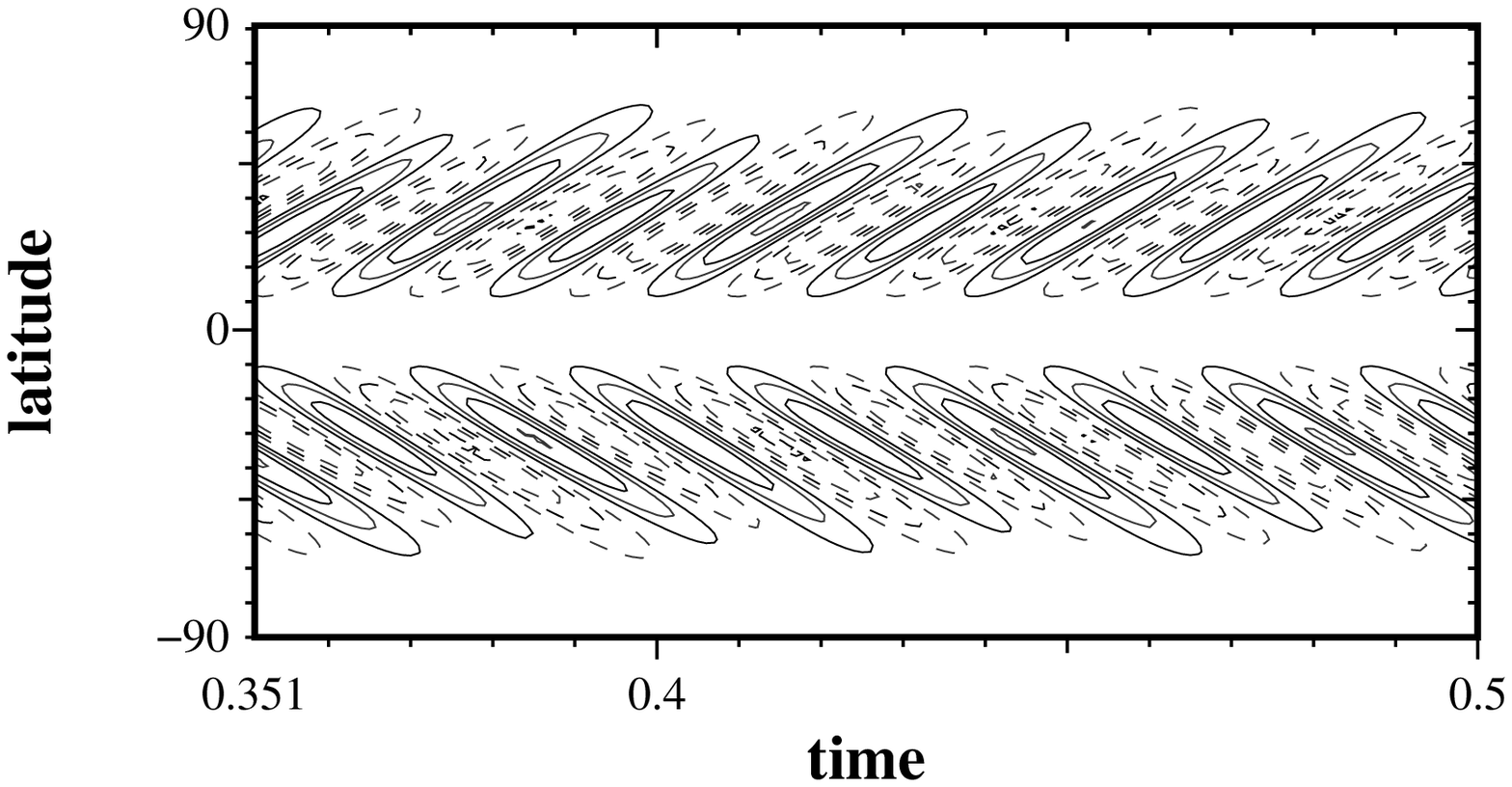}\cr
\includegraphics[width=7cm]{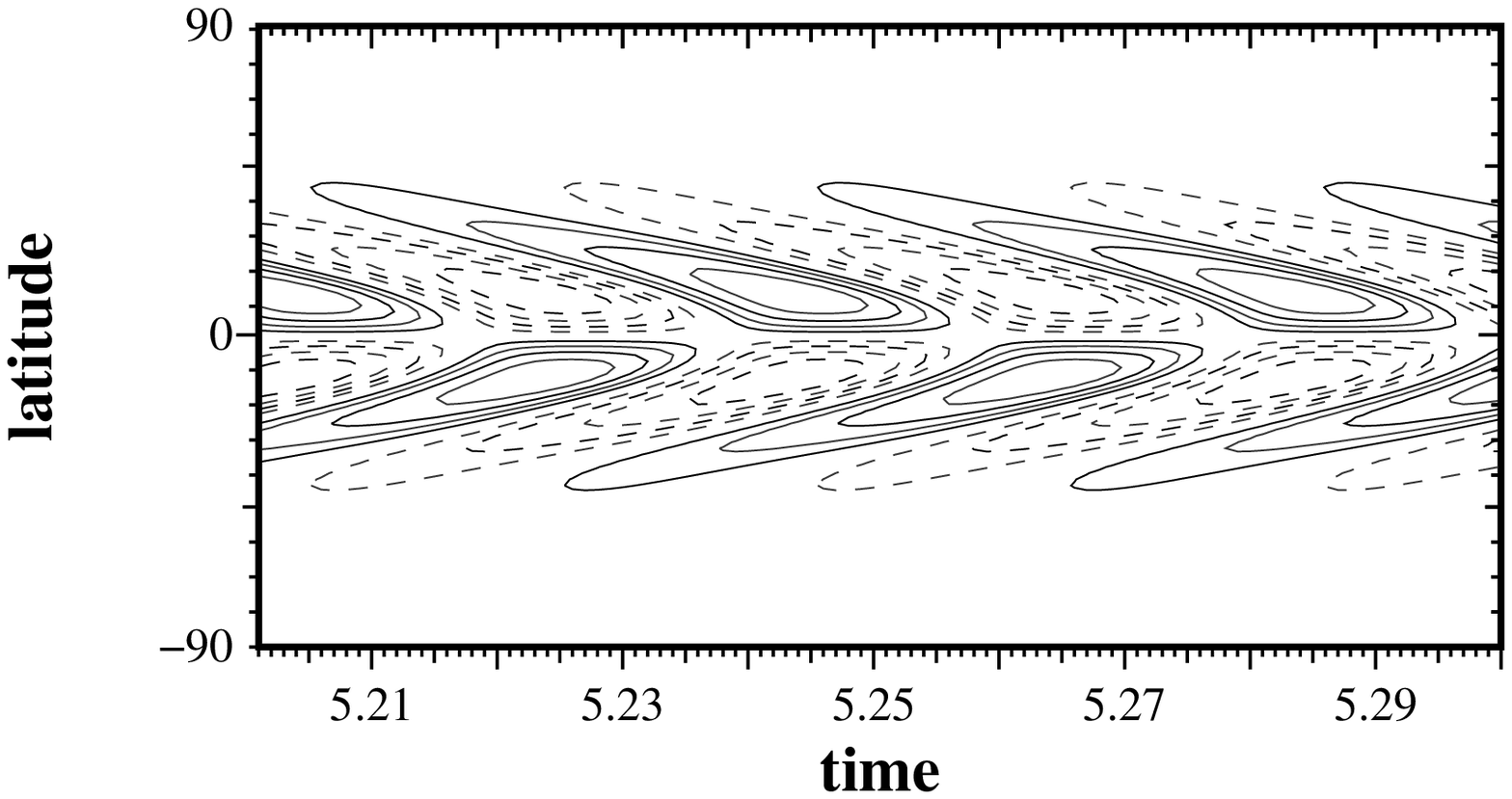} &
\includegraphics[width=7cm]{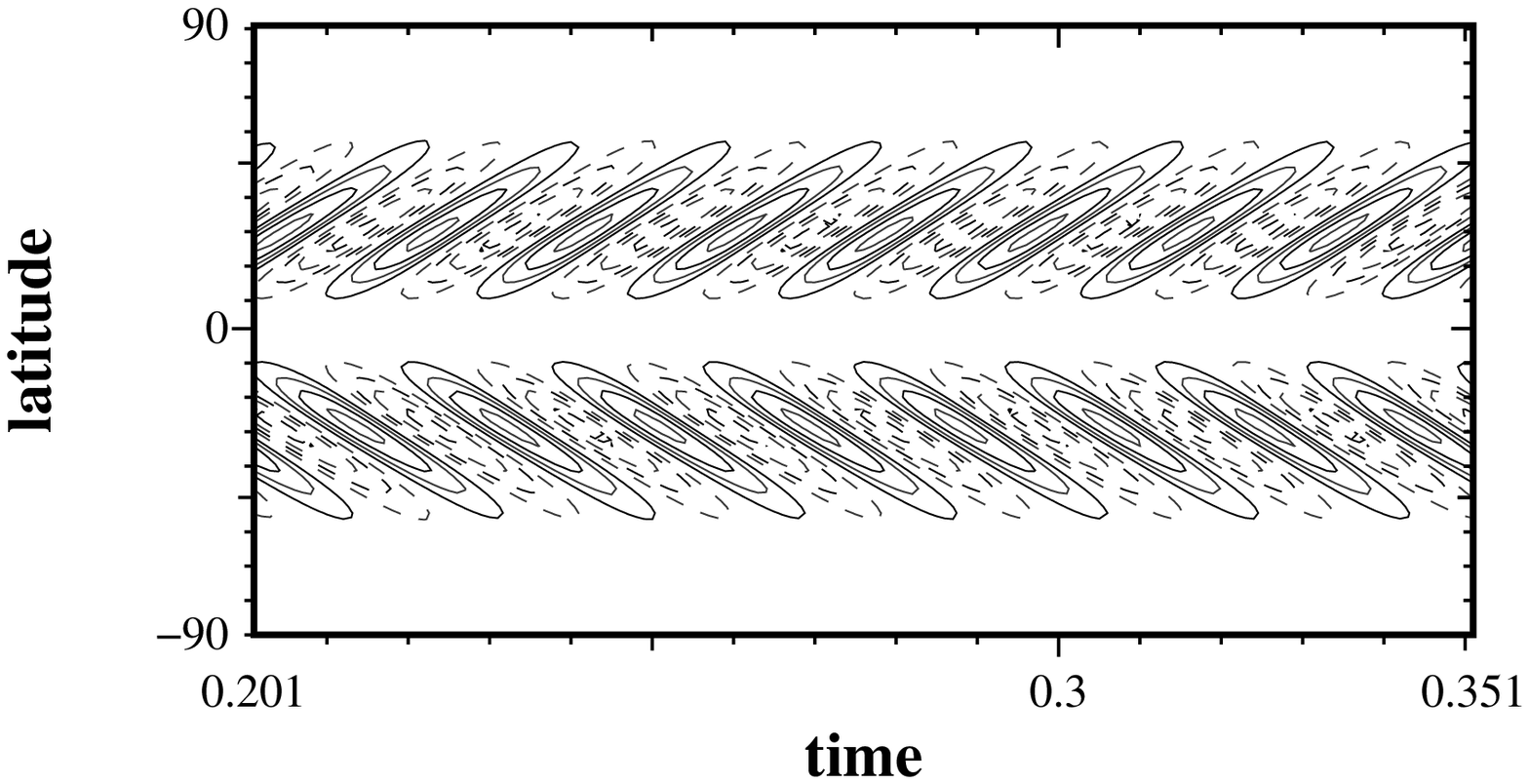}\cr
\end{tabular}
\end{center}
\caption[]{\label{butt1-19}  Butterfly diagrams for the sub-surface
field. Left hand column has $C_\alpha<0$, right hand column has
$C_\alpha>0$. Rows 1-4 have the quasi-solar rotation law, rows 5 and
6 the quasi-cylindrical law.  First row: Models 1 and 19 with
$i_\alpha=3, m=4$. Second row: Models 6 and 24 with $i_\alpha=1,
m=4$.  Third row: Models 8 and 26 with $i_\alpha=0, m=2$.  Fourth
row: Models 12 and 30 with $i_\alpha=1, m=2$. Fifth row: Models 1c
and 13c, with $i_\alpha=3, m=2$. Bottom row: Models 7c and 19c, with
$i_\alpha=3, m=4$. }

\end{figure*}

Only odd parity (dipole-like) solutions were studied. Models 1-18
have $C_\alpha<0$, the conventional fix to get near-surface fields
migrating in the solar sense. Models 19-30 (not all numbers present)
have $C_\alpha>0$. Latitude-time diagrams for the pairs (1,19),
(6,24), (8,26), and (12,30), which have the same values of the
parameters $i_\alpha$ and $m$ and the same $|C_\alpha|$, are shown in
the top four rows of Fig.~\ref{butt1-19}.
Here and below, "near-surface"
refers to a fractional radius of ca. $0.93$, and "deep" to radius ca.
$r = r_0$.

\begin{table*}
\caption{The general features of the solutions with solar-like
rotation law. EW, PW, SW denote equatorwards, polewards, standing wave
respectively for migration of $B_\phi$ in a near-surface layer.
Models 1-18 have $C_\alpha<0$, Models 19-30 have $C_\alpha>0$.
Models (1,19), (6,24), (8,26) and (12,30) have identical parameters,
except that the sign of $C_\alpha$ has been reversed, i.e. the pairs
have the same value of $|C_\alpha|$. }
\begin{tabular}{|c|c|c|l|}
\hline
Model & $i_\alpha$  & $m$  & Notes \\
\hline
1 &       3  & 4 & predominantly EW at low latitudes, weaker PW branch at high latitude\\
2 &       0  & 4 & rather strange butterfly, very weak EW at low lats, no migration at high lats\\
3 &       2  & 4 & low lat EW, no high lat feature \\
4 &       6  & 4 & steady solution\\
5 &       7  & 4 & steady solution\\
6 &       1  & 4 & predominantly EW at low latitudes, very weak PW branch at high latitude\\
7 &       3  & 2 & predominantly EW at low latitudes, weaker PW branch at high latitude\\
8 &       0  & 2 & strong PW branch at high lat, EW at low\\
9 &       2  & 2 & low lat EW, very weak high lat PW \\
10 &       6  & 2 & steady solution\\
11 &       7  & 2 & steady solution\\
12 &       1  & 2 & EW at low lats, weak PW at high lat \\
13 &       3  & 0 & weak EW at low latitudes, much stronger PW branch at high latitude\\
14 &       0  & 0 & strong PW at high latitudes, very weak EW branch at low latitude\\
15 &       2  & 0 & steady solution \\
16 &       6  & 0 & EW at very low lats, no high lat features\\
17 &       7  & 0 & steady solution\\
18 &       1  & 0 & strong PW at high lat, very weak EW at low lat. Solns steady at large $C_\alpha$ \\
19 &       3  & 4 & almost no migration - near SW low lats, with very weak PW drift \\
24 &       1  & 4 & much as Model 19 \\
26 &       0  & 2 & EW, extending over most latitudes \\
30 &       1  & 2 & mild EW (near SW), over most latitudes \\
\hline
\end{tabular}
\label{solarsummary}
\end{table*}
\subsubsection{Meridional circulation}
\label{mer_circ}

Here we investigate the effects of an arbitrarily imposed
 meridional circulation, in addition to solar-like
differential rotation and $\alpha$-effect, on the butterfly diagram.

\begin{figure}
\includegraphics[width=8cm]{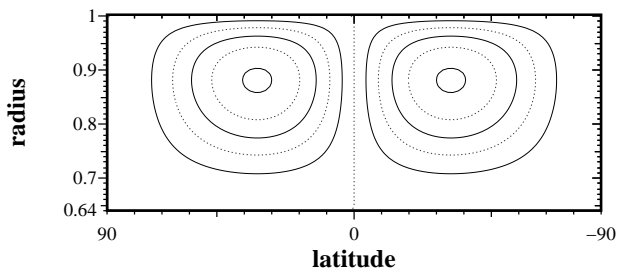} \\
\caption[]{\label{merid} Streamlines of the circulation defined by
(\protect\ref{defcirc}). With $Rm>0$ the surface flow is polewards. }
\end{figure}

We take a circulation determined by a stream function:
\begin{equation}
\psi= Rm\frac{1}{2}(r-r_0)^2(r-1)\sin^2\theta\cos\theta,
\end{equation}
so that
\begin{eqnarray}
\label{defcirc}
u_r&=&\frac{1}{r^2\sin\theta}\frac{\partial\psi}{\partial\theta}, \nonumber\\
u_\theta&=& -\frac{1}{r\sin\theta}\frac{\partial \psi}{\partial r},
\end{eqnarray}
where $r_0=0.64$ corresponds to the base of the dynamo region.
(Taken literally, in the solar case this implies the circulation
penetrating into an overshoot region, but limited experimentation
with the circulation restricted to $r>r_0=0.7$ suggests  little difference).
{Here $Rm=U_0R/\eta_0$, where $U_0$ is the maximum value of $u_\theta$ at
the surface.}
This circulation has a single cell in each hemisphere, with polewards flow
at the surface if $Rm>0$.
The streamlines $\psi={\rm const}$ are shown in Fig.~\ref{merid}. 

In our models, the $\alpha$ effects and the maximum of  the angular velocity shear are not spatially separated, as in, e.g., flux transport models based on the Babcock-Leighton paradigm (e.g., \cite{dikpaticharbonneau99}). Therefore, the meridional flow does not play the crucial role it has in those kinds of models, but it can still modify the migration of the activity waves in a given layer when its speed there becomes comparable to the migration speed of the wave in the absence of circulation which is mainly established by the product $|\alpha \partial \Omega / \partial r|$. 

\begin{figure}
\includegraphics[width=7cm]{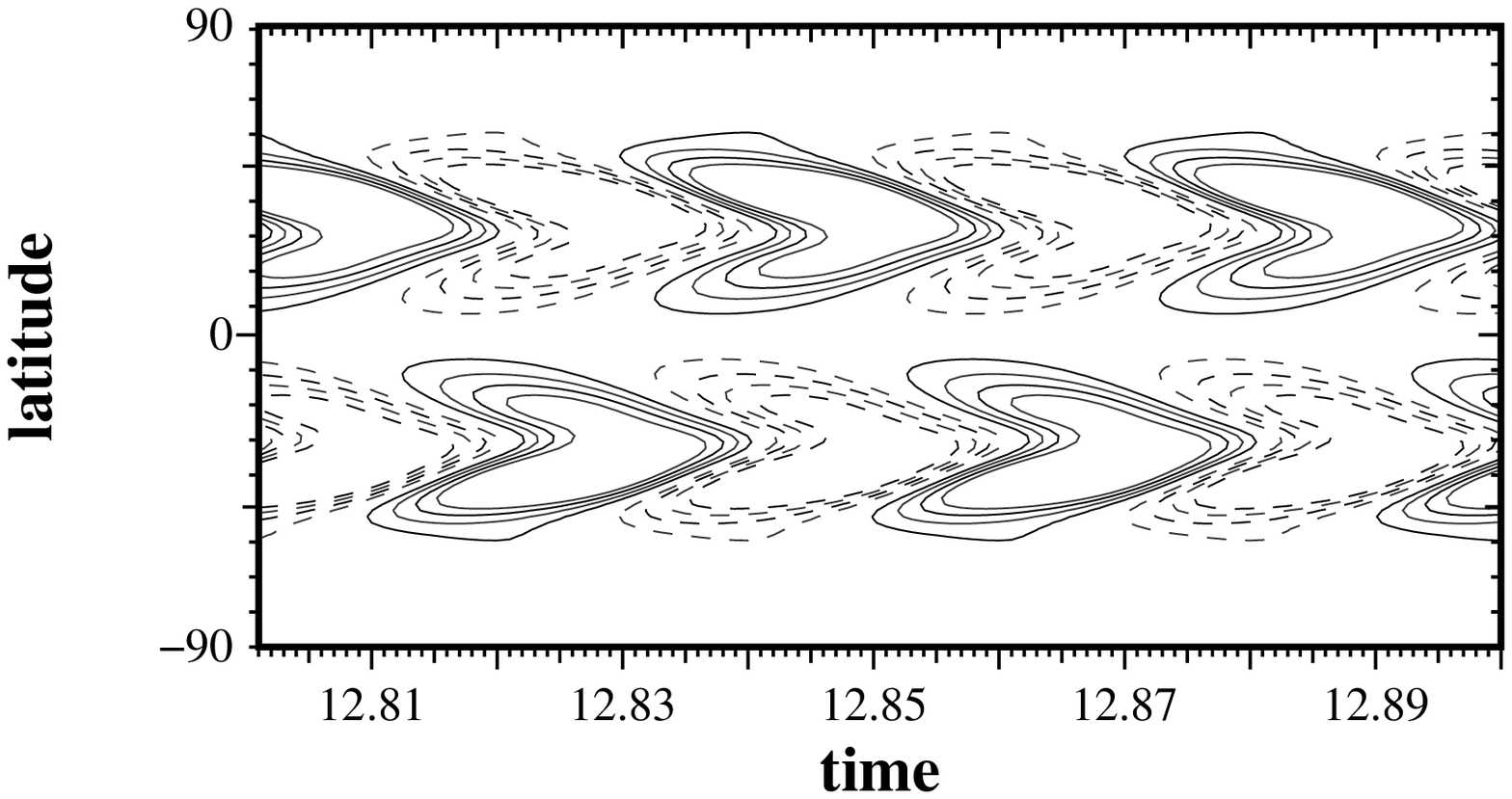} \\
\includegraphics[width=7cm]{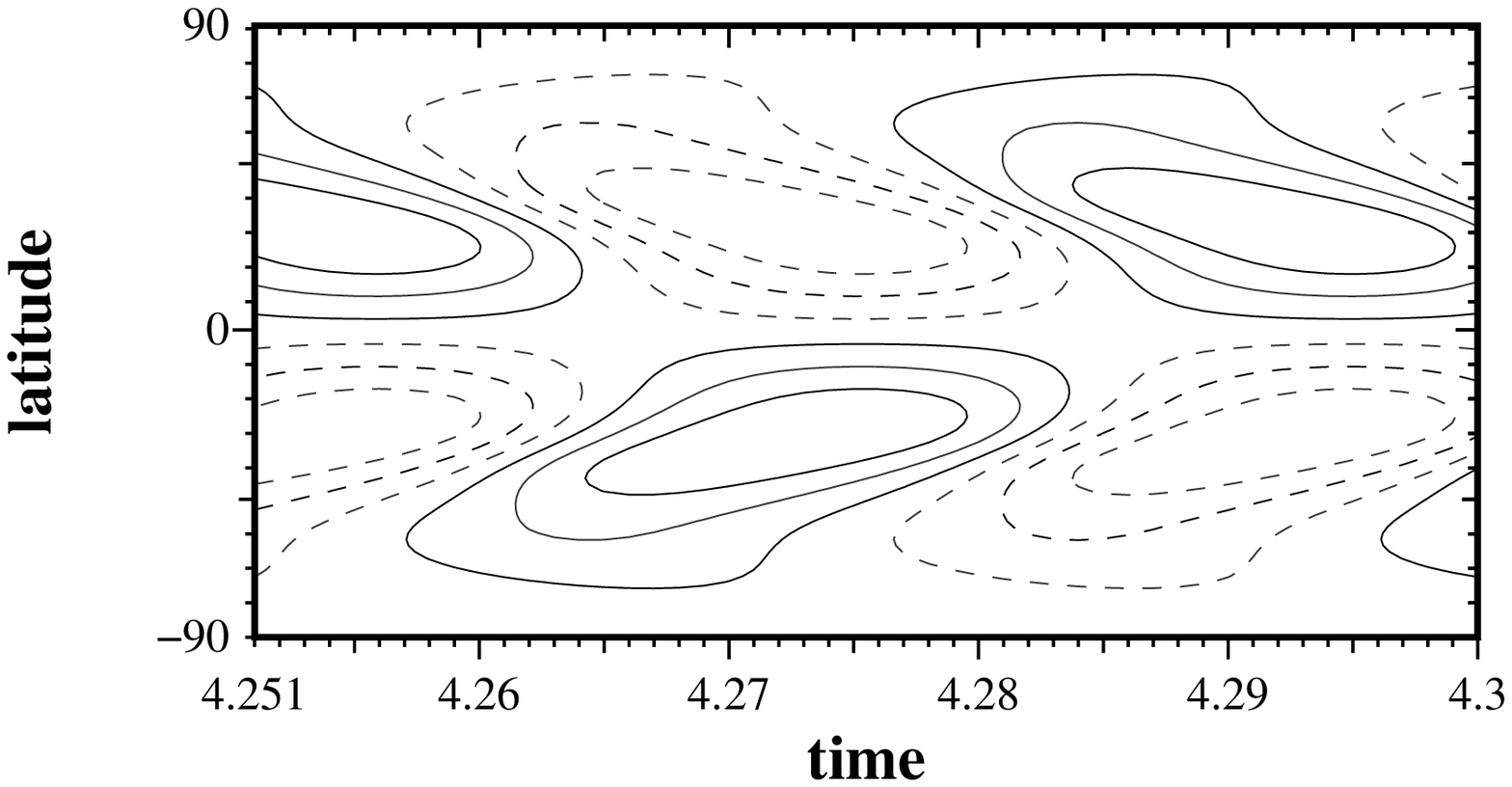}\\
\includegraphics[width=7cm]{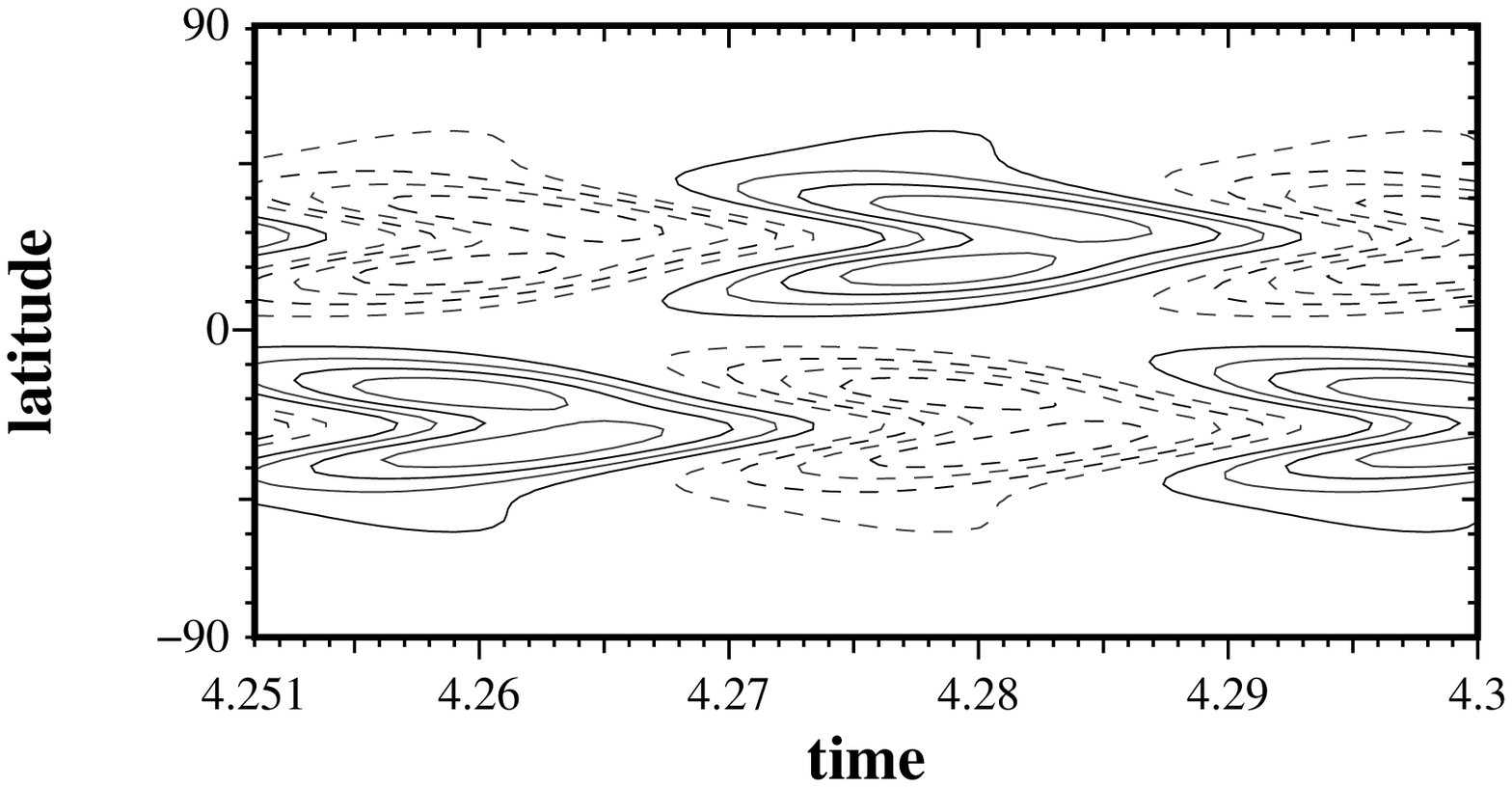}
\caption[]{\label{1099} 
Butterfly diagrams for Model~26 (solar rotation, $C_\alpha>0, i_\alpha=0, m=2$) : top - without meridional circulation ($Rm =0$), bottom of the convection zone; middle - $Rm = 20$, sub-surface;
bottom - $Rm = 20$, bottom of the convection zone. Solid/broken contours denote positive/negative
values of $B_\phi$ respectively.
}
\end{figure}

We first look at Model~26 (with a solar-like rotation). In the absence of
any meridional circulation, the near-surface migration is equatorwards
(Table~\ref{solarsummary}), but the deep butterfly diagram has polewards
and equatorwards branches, the polewards being somewhat stronger, see
Fig.~\ref{1099}, top panel. In the presence of our one-cell
circulation, polewards at the surface, the near-surface migration
remains equatorwards for $Rm\la 40$, but the deep equatorwards branch is strengthened,
see the lower panels of Fig.~\ref{1099}. For larger values of $Rm$,
the near-surface migration acquires a weak polewards branch when $Rm\ga
100$, but the deep migration develops two discrete patterns that are
close to standing waves.  Note also that, here and below, we
do not claim to have explored exhaustively the parameter space, but
just to have sampled a few, we hope fairly representative,
solutions. Other behaviours may well remain to be found.

\subsubsection{Quasi-cylindrical rotation law}

Here a quasi-cylindrical rotation law, as \cite{covasetal2005} and
shown in the right hand panel of Fig.~1, is used. We take $C_\alpha=-11.7,
C_\omega=1.3\times 10^5$, i.e. significantly supercritical values
that might correspond approximately to $\Omega=3\Omega_\odot$.
Exceptionally, Models 13 and 19 have $C_\alpha=+11.7>0$.
Results are summarized in Table~3 (note that these models are
labelled "1c" etc to distinguish them  from the models with a solar-type
rotation law). Butterfly diagrams for the sub-surface toroidal
field for the pairs of Models (1c, 13c), and (7c,  19c) are shown in
Fig.~\ref{butt1-19} in the fifth and sixth rows, respectively. The models in each of these pairs have
the same parameters, except that for 13c and 19c, $C_\alpha>0$.

\subsubsection{An intermediate rotation law}

\begin{figure}
\centerline{\includegraphics[width=4cm]{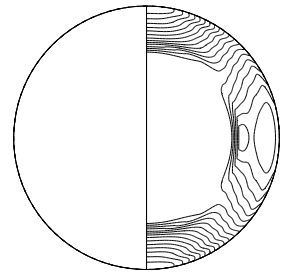}}
\caption{Synthetic rotation curve from combining solar-like and
quasi-cylindrical rotation laws with 50\% weighting.}
 \label{synth}
\end{figure}

\begin{figure}
\includegraphics[width=8cm]{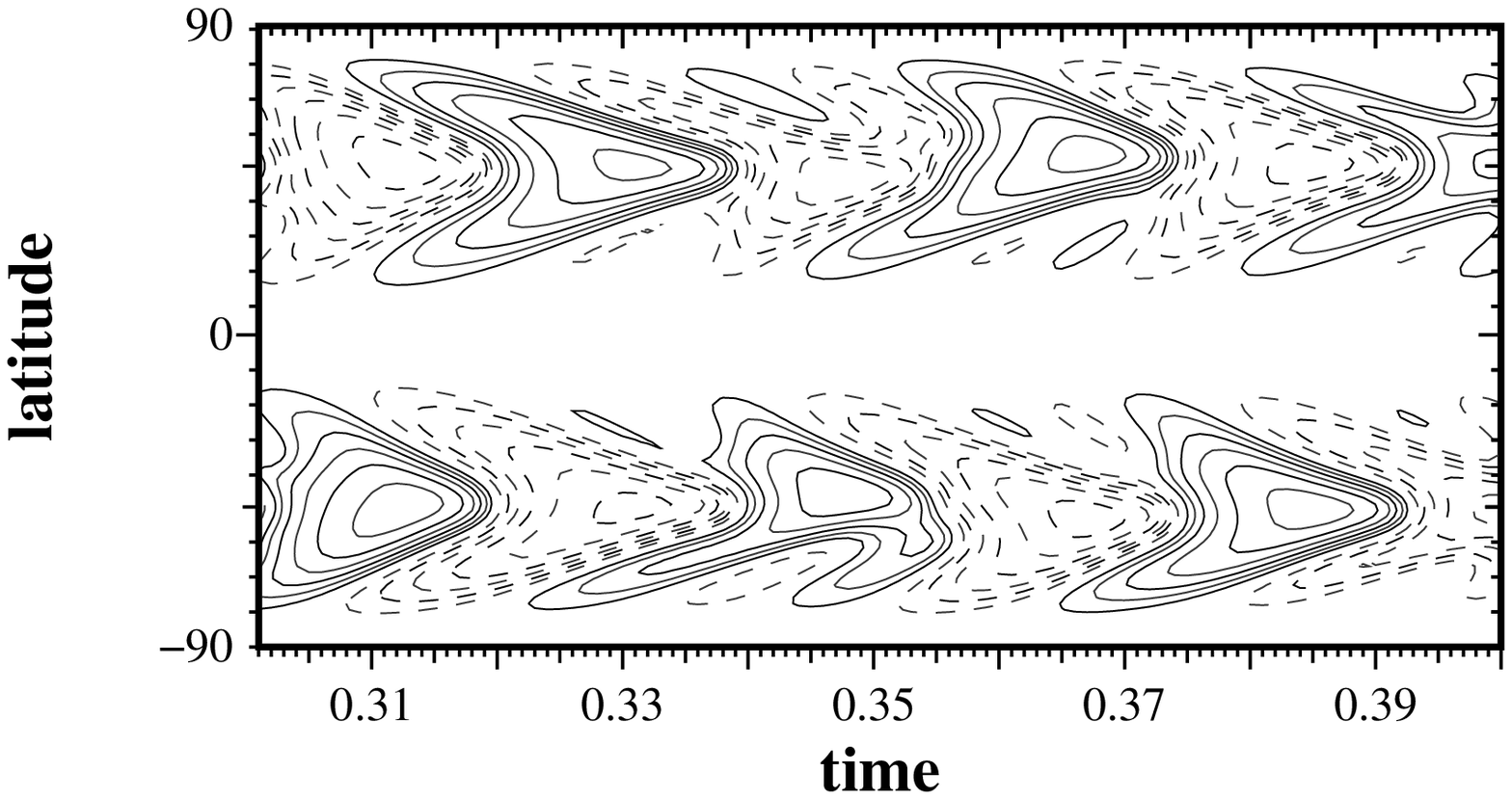} \\
\includegraphics[width=8cm]{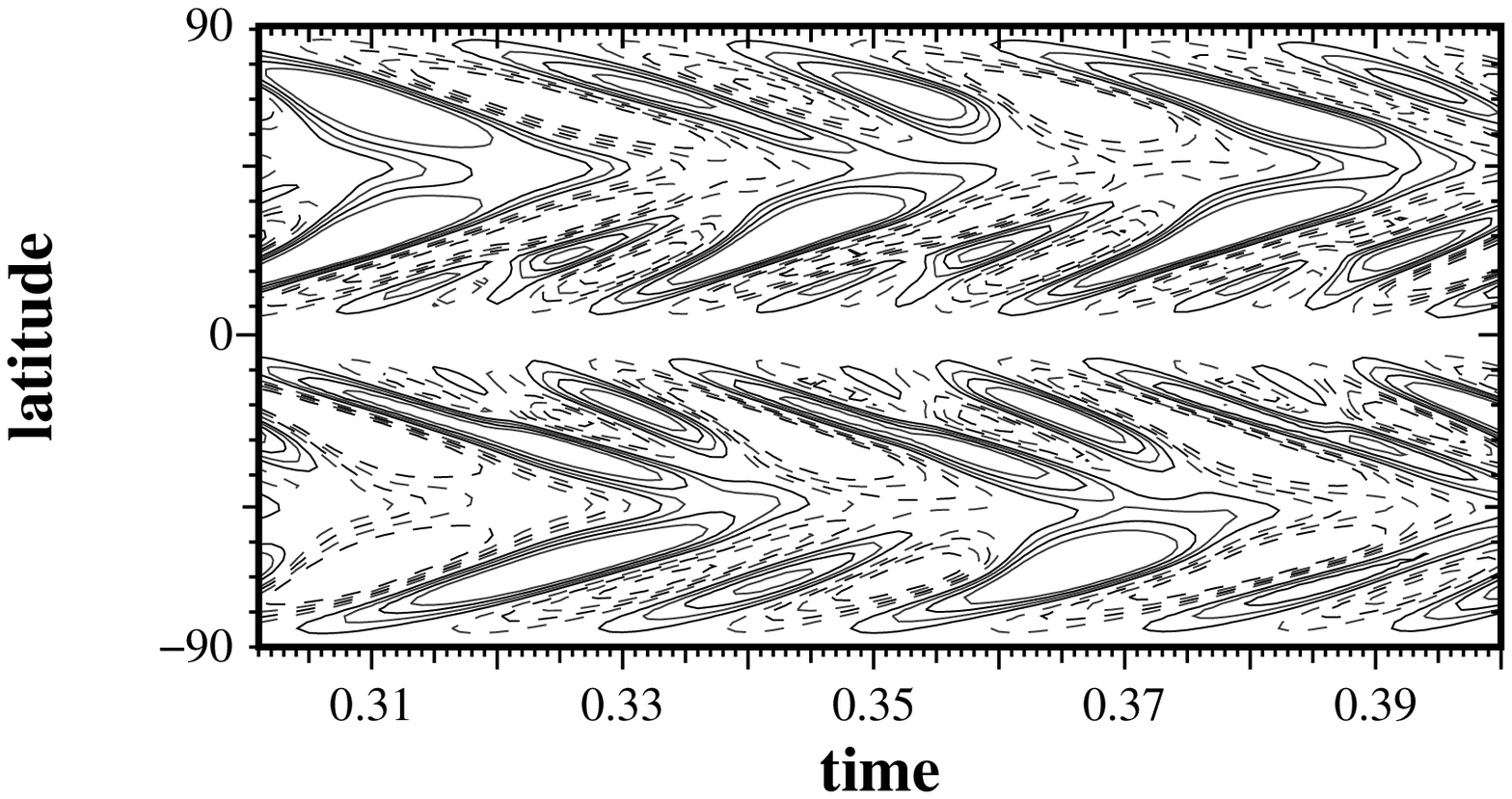}
\caption[]{\label{1099synth} Butterfly diagrams  for the synthetic
rotation law (combination of solar-like and quasi-cylindrical): top
panel - shallow,  lower panel - deep parts of the convective zone.}
\end{figure}

 Although several stars in our study are quite rapid rotators,
the cylindrical rotation law discussed in the previous section is not the
only possibility for their rotation.
It appears plausible that a regime intermediate
between the solar-like and the quasi-cylindrical rotation laws may be
more appropriate for such stars. Therefore, we explored the butterfly
diagram obtained by adopting such a synthetic rotation curve, i.e.,
mixing these laws with weights of about 50\% (Fig.~\ref{synth}).
Using this $\Omega(r, \theta)$, we obtained butterfly diagrams for
$B_\phi$ in both the deep and shallow parts of the convective zone.
These are
characterized by field migrating from low and high latitudes
towards some intermediate latitude, even without including any
meridional circulation (Fig.~\ref{1099synth}).
Neither the solar-like nor the quasi-cylindrical rotation law produced such a
behaviour.

\subsubsection{Models with a deep convection zone}
\label{HR1099}

Since $r_0=0.64$ probably gives a CZ that is too shallow
for HR~1099, we experimented with
models having a deeper dynamo zone ($r_0=0.2$), with both solar-like  and
quasi-cylindrical rotation laws and also their equally weighted combination.
The role of a simple meridional circulation, as parametrized
by $Rm$, was also considered (cf. Sect.~\ref{mer_circ}).

In general,  the deep convective zone appears  to favour
 steady solutions when $Rm
\ne 0$, or standing wave (SW) solutions. The latter may
partially be due to the restricted latitudinal extent at the
bottom of a deeper convective zone. This conclusion agrees with
the previous results of \cite{mossetal04}.

The solar rotation law  sometimes gives low latitude equatorwards and high
latitude polewards branches simultaneously present near the surface,
similar to the models for solar-like stars with $r_0=0.64$
previously discussed.  With $C_\alpha<0$, with the quasi-cylindrical
rotation law, deep and surface migration was either SW or equatorwards
depending on the sign of $Rm$, and becomes steady for large enough
$|Rm|$. If $C_\alpha>0$, then solutions with polewards surface migration
and SW or vacillatory (V) behaviour deep down were found for $Rm>0$.
With $Rm<0$, for the same $C_\alpha$ surface migration patterns were
a combination of SW and equatorwards, again with SW in the deep CZ. Again
solutions are steady for large enough $|Rm|$. The richest choice of
possibilities was obtained for the intermediate rotation law with
$C_\alpha$ and $Rm$ varying (Table~\ref{deep}). It is notable that
examples with surface polewards migration were not found.
{ These investigations clearly demonstrate that the effects of
advection can dominate the basic dynamo wave when $Rm$ is significantly
non-zero.}
We conclude that none of these butterfly diagrams can reproduce the
phenomenology observed in HR~1099 (case I of
Sect.~\ref{observations}).

\begin{figure}
\includegraphics[width=7cm]{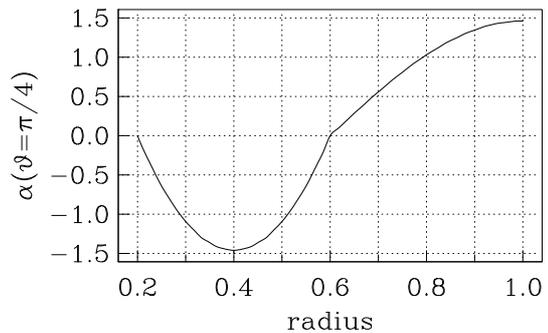}\\
\caption{Dependence of $\alpha$ on radius at $\theta=\pi/4$ for the
case where $\alpha$ changes sign with radius.} \label{ialpeq18}
\end{figure}

\begin{figure}
\includegraphics[width=8cm]{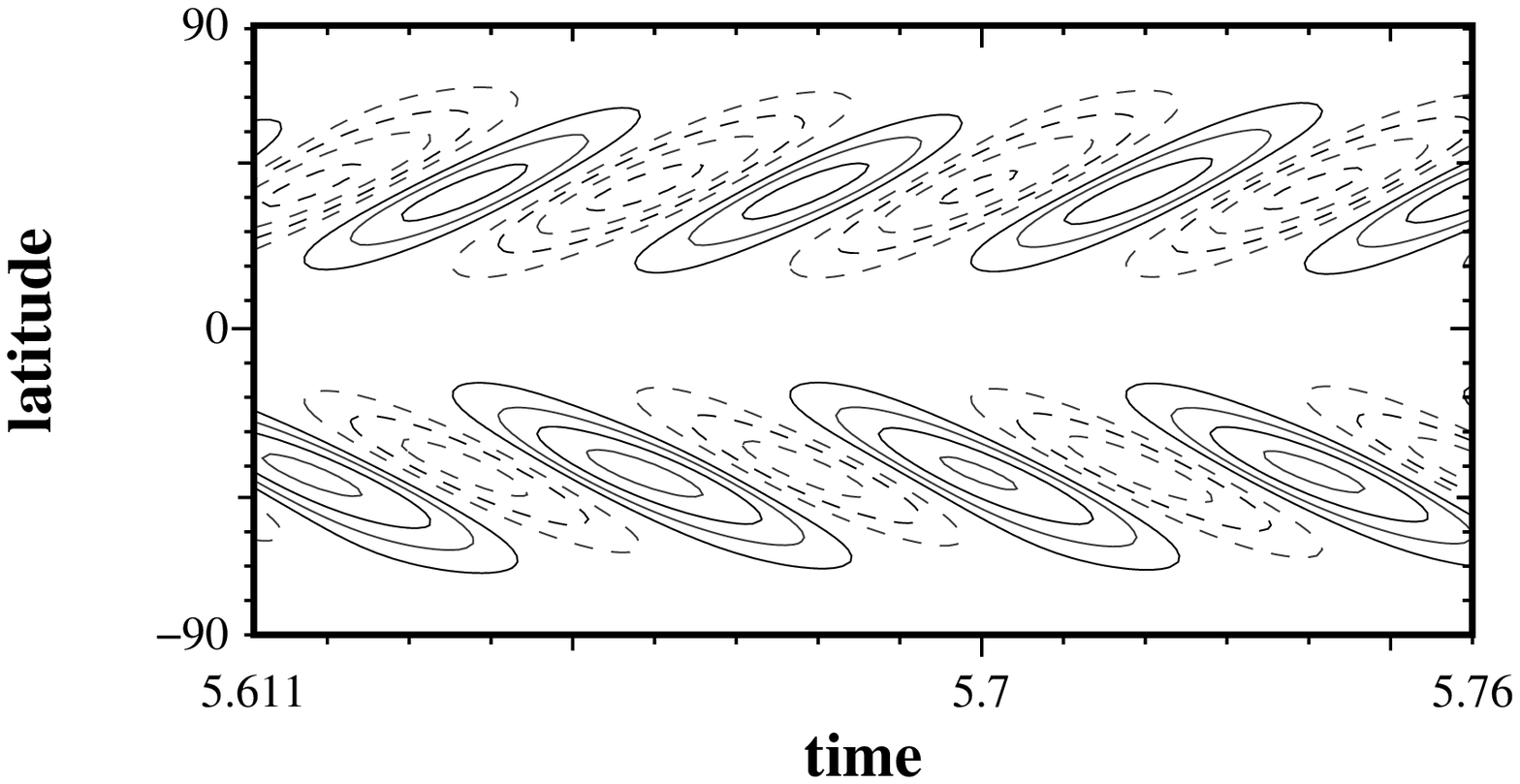} \\
\includegraphics[width=8cm]{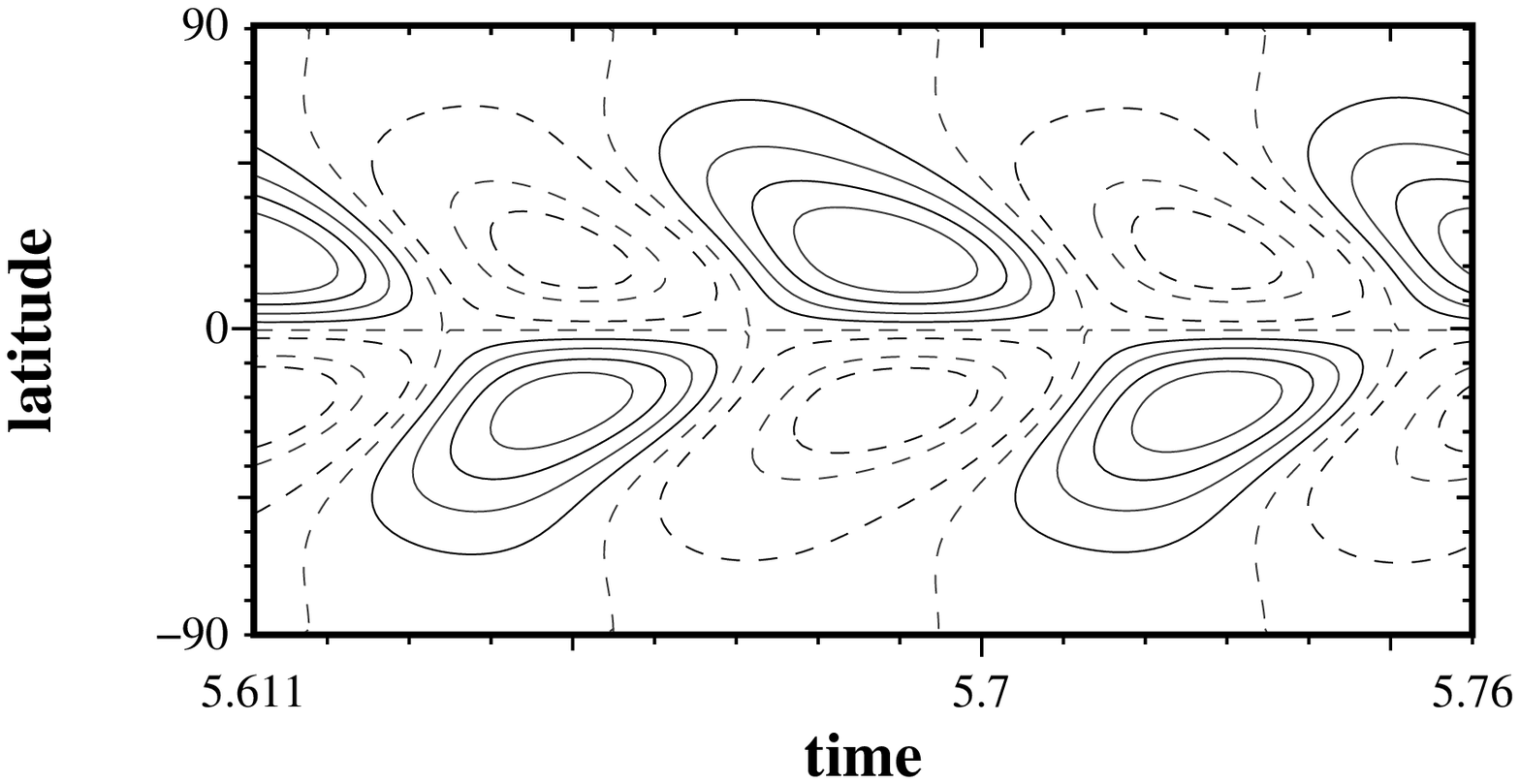}
\caption[]{\label{deepCZ}  Butterfly diagrams resembling that of
HR~1099 for a deep convective zone with a change of sign for
$\alpha$ \protect(Fig.~\ref{ialpeq18}), quasi-cylindrical rotation,
$i_\alpha=0$:  sub-surface (upper panel) and deep (bottom panel).  }
\end{figure}

A more promising result (Fig.~\ref{deepCZ}) can be obtained when using an $\alpha$
profile which changes sign with depth { (see Fig.~\ref{ialpeq18})  }
which gives clearly opposed
waves near top and bottom of the convective zones\footnote{Turbulent convection simulations by \cite{Kapylaetal09} do
indeed show an $\alpha$ effect that changes sign from the bottom to
the top of the convective domain (cf. the sign of their tensor
coefficient $\alpha_{\rm yy}$, that corresponds to the scalar coefficient
$\alpha$ of our simplified mean-field model, in their Figs. 3 and
12). { This effect also appears when $\alpha$ is determined from
some turbulence models -- see e.g. \cite{br} and references therein.} }.
 In this case, we only investigated the quasi-cylindrical rotation law
for a very limited range of dynamo parameters. This result is
consistent with the results of \cite{ms07} concerning dynamo waves
propagating in two separate layers, and allows the behaviour sought
with appropriate choices of $C_\alpha$ in the two layers. The
separation of the layers, that was assumed by \cite{ms07} arbitrarily,
may be achieved here by exploiting the freedom given by the depth of
the convective zone. Of course, this assumes that the deep and
surface magnetic fields somehow jointly contribute to the surface
activity manifestations (see Sects.~\ref{comparison}
and~\ref{concl}).

\begin{table*}
\caption{The general features of the solutions with the quasi-cylindrical
rotation law. EW, PW denote equatorwards, polewards respectively for
$B_\phi$ migration in a near-surface layer. Note $C_\alpha=+11.7>0$ in
Models 13 and 19.}

\begin{tabular}{|c|c|c|l|}
\hline
Model & $i_\alpha$  & $m$  & Notes \\
\hline
1c &       3  & 2 & strong EW at low latitudes, no high latitude features\\
2c &       0  & 2 & strong EW at low latitudes, no high latitude features\\
3c &       2  & 2 & strong EW at low latitudes, no high latitude features \\
4c &       6  & 2 & strong EW at low latitudes (slightly more extended than\\
  &          &   &   above), no high latitude features\\
5c &       7  & 2 & similar to Model 4c\\
6c &       1  & 2 & similar to Model 4c\\
7c &       3  & 4 & similar to Model 1c\\
8c &       0  & 4 & similar to Model 2c \\
9c &       2  & 4 & again, low lat EW\\
10c &       6  & 4 & similar to Model 9c \\
11c &       7  & 4 &  as Model 10c\\
12c &       1  & 4 &  as Model 10c\\
13c &       3  & 2 &  $C_\alpha>0$. PW at mid-latitudes\\
19c &       3  & 4 &  $C_\alpha>0$. PW at low latitudes\\
\hline
\end{tabular}
\label{cylsummary}
\end{table*}

\section{Comparison of the 2D dynamo models with the observations}
\label{comparison}

With  the solar rotation law, details of migratory patterns are sensitive
to the spatial dependence of $\alpha$, and even the general rule that the
direction of migration is governed by the sign of $C_{\alpha} C_{\omega}$
appears not always to hold true.
The most striking case is that of Models~8 and 26 -- see the third 
row of Fig.~\ref{butt1-19}. Model~8 (left panel) has $C_{\alpha} < 0$ and shows a
 strong polewards branch in addition to an equatorwards low-latitude branch. 
Reversing the sign of $C_{\alpha}$ but maintaining the same rotation law and spatial structure of the $\alpha$ effect, we see only a pronounced equatorwards branch, contrary to simple intuitive expectation. The effect of switching from $C_{\alpha} < 0$ to $C_{\alpha}>0$ is also remarkable and not simply intuitive for Models 6 and 24 (on the second row of Fig.~\ref{butt1-19}) and for Models 12 and 30 (on the fourth row); these differ in the latitudinal localization of 
the $\alpha$ effect ($m=4$ and $m=2$, respectively).

With the quasi-cylindrical rotation law, migratory patterns are quite
insensitive to the exact form of $\alpha$, the only significant
change occurs when
the sign of $C_{\alpha} C_{\omega}$ is reversed. The rule linking
the sign of $C_{\alpha} C_{\omega}$ to the migration direction appears to hold.
Plausibly this is because
the spatial structure of $\Omega$ is much simpler than in the solar-like case,
and $\alpha$ cannot be 'tweaked' so as to give extra weight to
regions of the envelope with anomalous gradients of $\Omega$, unlike
in the solar case.

Our general impression from the above analysis can be summarized as
follows. Dynamo models with $r_{0}=0.64$ and a fixed sign of
$\alpha$ through the CZ can provide polewards migrating patterns, and  observers
understand  activity in some stars as a manifestation of a polewards
propagating pattern (see cases III and IV in
Sect.~\ref{observations}). Models with a solar-like rotation law tend
to show both equatorwards and polewards branches, the latter with an
intensity that depends on  the spatial distribution of the $\alpha$
effect. It is possible to reproduce  case IV by, e.g., Model 8,
while the case with a stronger equatorwards  and a weaker polewards
branch, reminissent of the behaviour observed in the Sun, can be
compared with Model 6 (see Fig.~\ref{butt1-19}, left hand panel in the 
second row). The case with polewards migration only, i.e.
case III of our classification, is unlikely to be reproduced with a
solar-like rotation law, but can be reproduced with a cylindrical
law (cf. e.g. Model 13c with $C_{\alpha} > 0$,  the right hand panel, 
fifth row of Fig.~\ref{butt1-19}). This rotation law
may be characteristic of stars rotating significantly faster than
the Sun (see, e.g., \cite{lanza05}).

On the other hand, a  pattern of spots  extending from the high to
the low latitudes without any clear evidence of migration during the
activity cycle (case II in Sect.~\ref{observations}), may  be
interpreted by, e.g. Model 24  (see Fig.~\ref{butt1-19}, right hand
panel in the second row). That model is characterized by a
stationary pattern spanning a latitude range from the equator up to
$\sim 65^{\circ}-70^{\circ}$. Also Model 19  (Fig.~\ref{butt1-19}, right 
hand panel, first row) shows a similar
stationary pattern, but more localized in latitude. Model~12 
(Fig.~\ref{butt1-19}, left hand panel of the fourth row) shows 
a  weak polewards branch that, although probably not capable of producing 
photospheric spots, could be detectable through the chromospheric 
Ca~II H\&K flux modulation. Therefore, it might correspond to our case~VI
because the pattern localized in latitude gives rise to an almost
constant primary periodicity in the chromospheric flux modulation,
while the migrating branch produces a secondary periodicity that
varies along the activity cycle.

When interpreting the observations, one should consider that 
the outer contours of the model butterfly diagrams can be affected by the turbulent diffusion.  
Nevertheless, the direction of migration, which is the relevant information for our study, is not
modified, as can immediately be seen by considering the direction of
migration given by the interior contours of the diagrams that correspond
to increasingly stronger fields. In some of the diagrams, such as those
of Model 6 (on the left second row in Fig.~\ref{butt1-19}) or Model 12 (on the left
fourth row) we plot only one contour to represent the polewards branch,
but the reality of that branch is confirmed 
by more detailed investigation. (A single contour merely indicates that the
field strength in that branch is relatively small.)
Moreover, turbulent diffusion in our model is uniform  in 
the outer layers of the convection zone which implies that it cannot
favour a specific direction of migration of the field, i.e. it does not
introduce any preference for equatorwards or polewards motion of
butterfly contours.

In the case of HR~1099 (case I of Sect.~\ref{observations}), the
point is that the image of polewards patterns extracted from the
observations is quite different from that emerging from the theory in
all the models with $r_{0}=0.64$ introduced in
Sect.~\ref{2d_dynamos}, irrespective of the adopted rotation law or
the inclusion of a meridional circulation. We recall that HR~1099
displays two patterns that migrate in opposite directions, and the
migration  that begins nearer to the equator is polewards.

If there is simultaneous migration polewards and equatorwards through
the same latitudes, then it is difficult to see how any simple
mean-field model can reproduce it. We
attempt to interpret the two oppositely propagating activity
patterns as originating from different spatial volumes (see
Sect.~\ref{HR1099}). Specifically, in the case of HR~1099, we need
to consider an $\alpha$ effect that changes sign with radius or the
presence of two distinct dynamo layers with opposite signs of
$\alpha$. The main difference with respect to the models with
$r_{0}=0.64$ is that now we consider the contribution to surface
activity from {\it both} the deep and the shallow dynamo layers,
while in the other cases we assumed that the field pattern at the
top of a single dynamo layer directly accounts for the observed
butterfly diagram.

We recognize that there are unresolved difficulties with this idea. 
For example, there is the problem of storing toroidal fields in relatively shallow superadiabatic layers
without a too rapid loss through buoyancy instabilities.
This issue has been addressed, {\it inter alia}, by Brandenburg~(2005, 2009). 
{ We  note in passing that in our model the toroidal field strength
in the upper region is about 20\% of that in the lower.
This might assist downward turbulent pumping to stabilize the field by
its reduced magnetic buoyancy force (proportional to $B^{2}$), 
possibly acting preferentially at the base of a stellar supergranular layer 
analogous to the solar supergranulation (but at a somewhat greater depth 
because the convection in a subgiant is expected to have larger vertical scales).}
Turning back to the Sun, other difficulties for a dynamo model operating in the subsurface shear 
layer  may include an incorrect phase difference between poloidal 
and toroidal fields and a too weak  toroidal field due to the strong 
turbulent diffusion expected in those highly superadiabatic layers, cf. \cite{Dikpatietal02}. 

{ Nevertheless, we can attempt to reverse the argument: from both general considerations 
about mean-field models, and our particular simulations, we  are unable to
identify a mechanism to produce  the simultaneous presence of polewards 
and equatorwards  migration co-located in latitude, other than that discussed 
above. Thus we conclude that either this or a related mechanism does indeed
operate, or that the phenomenon is beyond the scope of mean-field theory.} 

Note that \cite{BerdyuginaHenry07} stress the nonaxisymmetric
behaviour of HR~1099 and other active stars, whereas we restrict
ourselves to axisymmetric models. We fully realize that departures
from axisymmetry may play a role in the phenomena displayed by HR
1099,  cf. e.g. \cite{MPS}, and seem essential to explain the
'flip-flop' phenomenon, e.g. \cite{Moss04}.
However
nonaxisymmetric dynamo models contain many additional uncertainties
and the main features of the observations seem to be reproduced by a
simpler model. 

\begin{table*}
\caption{ The variety of butterfly diagrams obtained with the
intermediate rotation law and the deep convective zone ($r_0=0.2$). Some
intermediate cases, not tabulated, were also found.
These behaviours were each found for a variety of parameters and, given
the fragmented nature of the parameter ranges investigated, we have  not here
specified parameters, and rather just indicate the possible range of behaviours.}

\begin{tabular}{|l|l||l|l|}
\hline
\hline
near surface  &   deep  & near surface & deep\\
\hline weak PW at mid-latitudes,   & SW &    V/SW near poles only & V/confused \\
SW at high lats &       & &\\
\hline
 near SW & weak EW & SW, concentrated at high lats & weak PW, mid-lats\\
 \hline
EW & almost no field present & EW & PW\\
\hline
 steady & steady & steady & PW\\
 \hline
EW & EW & PW & EW\\
\hline confused butterfly, no migration & near SW & EW & SW\\
\hline
V/confused, near pole only & V/confused & EW and SW & EW and SW\\
 \hline
\end{tabular}
\label{deep}
\end{table*}

\section{Discussion and conclusions}
\label{concl}

Our general conclusion is that it is becoming
realistic  to  construct a
framework for classification of { the very varied} 
stellar dynamo wave behaviours, and to relate this
to physical parameters of stellar convective zones. We have
presented above a very crude and preliminary outline for such a
template.

We confirm that the sign of $D = C_{\alpha} C_{\omega} \sim \alpha
\partial\Omega /\partial r$ is the main quantity which determines
the direction of activity wave propagation, { even though
considerable finer detail can be found in the results, e.g. in the
form of two branches at low and high latitudes, or of
standing wave patterns (cf. Sect.~\ref{comparison} and
Fig.~\ref{butt1-19}). However,} quite unexpectedly, we conclude from
the above plots that even if $D>0$ it is quite difficult to excite a
pronounced single polewards branch in stars with a solar-like
rotation law.  It follows that straightforward considerations
based on the sign of dynamo number $D$ are inadequate to
understand the occurence of polewards branches, and a 
careful examination of 2D models is important.  In our investigation, we consider only the direction of dynamo wave migration and  do not attempt to fit the observed spot latitudes.
The direction of migration (or standing wave behaviour) appears
to be a robust result that does not depend on the details
of the butterfly wings (which may be influenced by physical processes
that we have not considered).

Observations give a hint that one hemisphere of a star can contain two oppositely propagating
activity waves which are pronounced enough to be observable.
We find that, at least for some stars, rotation curves and
spatial distribution of the other dynamo governing quantities should
produce two activity patterns in a hemisphere, of more or less
comparable intensity. A phase difference between the equatorwards and the
polewards dynamo waves can exist. This idea emerges from looking at
the butterfly diagram for Model~1 (top left panel of Fig.~\ref{butt1-19}),
where there is a difference in the field intensity of the two
branches as well as a phase shift between the epochs when the
low-latitude branch reaches the equator and the high-latitude branch
reaches the highest latitude.

The fundamental result of our investigation is that the spot
migration  observed in main-sequence stars or the presence of
standing activity  waves can be explained by considering the time
evolution of the magnetic field at the upper boundary of a dynamo
shell, { even in a very simple dynamo model.  The various} behaviours can be reproduced  by changing the
spatial structure of the $\alpha$-effect and/or the internal
rotation law. Also a meridional flow may play a  role.

The other important point is that the behaviour of HR~1099 cannot be
explained by our simple one-layer models and the inclusion of a
meridional circulation does not change this conclusion.
On the other hand, a two-layer model, somewhat similar to that introduced by
\cite{ms07}, can explain the  behaviour of HR~1099 if the two spots
migrating in opposite directions are the result of magnetic flux
tubes originating in the upper and deeper dynamo layers,
respectively. However, this interpretation requires a significant
modification of the paradigm used for main-sequence stars. In other
words, while for main-sequence stars we use a single-layer dynamo
and assume that the observed spots correspond to the field at the
upper boundary of the dynamo domain, for the subgiant in HR~1099
(and for subgiant stars in general) we must assume the presence of
two dynamo layers separated by an inactive shell, with both layers
contributing to the observed spots at the surface.

From the viewpoint of dynamo theory, we have shown that by
exploiting the considerable range of freedom in choosing ill-known
physical quantities, we can produce a wider range of activity wave
behaviour than had previously been realized -- indeed  we can find
something approaching the observed range of stellar surface activity
waves. Of course, theory cannot at present say which of the models
(if any) correspond even loosely to reality. We just make the point
that the various observed behaviours are not incompatible with even
simple dynamo models. { The inherent uncertainties of mean field
models make it difficult to make a stronger or more useful
statement than this.}
 It remains to be clarified, for example,
whether there is a correlation of the dynamo wave behaviour with the
absolute value of the dynamo number, as suggested by our ordering of
the cases in Sect.~\ref{observations} by increasing values of $Ro$.
To take just one point considering a solar-like rotation, when the dynamo number is large (and $Ro$ is
small)  we generally see two waves with comparable magnetic field
strength, starting from mid latitudes and migrating towards the pole
and the equator,  respectively. On the other hand, when  the dynamo number is
small  (and $Ro$ is large) the equatorwards wave has a stronger field
than the polar wave.  { Further observational
evidence and characterization of polewards waves would be of special interest and value.}
Also the simultaneous presence of a standing dynamo wave at a fixed latitude
together with a migrating wave is of interest in interpreting the
behaviour of some of the Mt. Wilson stars.

Finally, we have of necessity based this exploratory paper on a
rather small number of relatively well observed stars (and of a
small subset of all possible dynamo models). We thus make the usual
plea for more, reliable, data in order to substantiate (or disprove)
our attempts at systematization.

\acknowledgements The authors are grateful to an anonymous referee for 
a careful reading and several useful comments on an earlier version of the manuscript. 
D.S. acknowledges financial support from RFBR
project 09-05-00076.


\end{document}